\documentclass[aps,pre,superscriptaddress,twocolumn,tightenlines,showpacs,floatfix,amsmath,amssymb,pdftex,usenames,dvipsnames]{revtex4-1}

\usepackage{amsmath}
\usepackage{amssymb}
\usepackage{bm}
\usepackage{bbm}
\usepackage{float}
\usepackage{tikz} 
\usepackage{graphicx}
\usepackage{subfigure}
\usepackage{multirow}
\usepackage{booktabs}
\usepackage[normalem]{ulem}
\def\id{\mathbbm{1}}
\def\integers{\mathbb{Z}}
\newcommand{\corr}[1]{\langle #1\rangle}
\newcommand{\vek}[1]{\mathbf{#1}}

\newcommand{\josko}[1]{{\color{black}#1}}
\newcommand{\jaakko}[1]{{\color{black}#1}}
\newcommand{\ke}[1]{{\color{black}#1}}

\begin{document}
\title{Hierarchy of orientational phases and axial anisotropies in the gauge theoretical description of generalized nematics}
\date{\today}

\author{Ke Liu}
\affiliation{Instituut-Lorentz for Theoretical Physics, Universiteit Leiden,
PO Box 9506, NL-2300 RA Leiden, The Netherlands}
\author{Jaakko Nissinen}
\affiliation{Instituut-Lorentz for Theoretical Physics, Universiteit Leiden,
PO Box 9506, NL-2300 RA Leiden, The Netherlands}
\author{Josko de Boer}
\affiliation{Instituut-Lorentz for Theoretical Physics, Universiteit Leiden,
PO Box 9506, NL-2300 RA Leiden, The Netherlands}
\author{Robert-Jan Slager}
\affiliation{Instituut-Lorentz for Theoretical Physics, Universiteit Leiden,
PO Box 9506, NL-2300 RA Leiden, The Netherlands}
\author{Jan Zaanen}
\affiliation{Instituut-Lorentz for Theoretical Physics, Universiteit Leiden,
PO Box 9506, NL-2300 RA Leiden, The Netherlands}

\begin{abstract}
The paradigm of spontaneous symmetry breaking encompasses the breaking of the rotational symmetries $O(3)$ of isotropic space to a discrete subgroup, i.e. a three-dimensional point group. The subgroups form a rich hierarchy and allow for many different phases of matter with orientational order. Such spontaneous symmetry breaking occurs in nematic liquid crystals and a highlight of such anisotropic liquids are the uniaxial and biaxial nematics. Generalizing the familiar uniaxial and biaxial nematics to phases characterized by an  arbitrary point group symmetry, referred to as \emph{generalized nematics}, leads to a large hierarchy of phases and possible orientational phase transitions. We discuss how a particular class of nematic phase transitions related to axial point groups can be efficiently captured within a recently proposed gauge theoretical formulation of generalized nematics [K. Liu, J. Nissinen, R.-J. Slager, K. Wu, J. Zaanen, Phys. Rev. X {\bf 6}, 041025 (2016)]. These transitions can be introduced in the model by considering anisotropic couplings that do not break any additional symmetries.  By and large this generalizes the well-known uniaxial-biaxial nematic phase transition to any arbitrary axial point group in three dimensions. We find in particular that the generalized axial transitions are distinguished by two types of phase diagrams with intermediate vestigial orientational phases and that the window of the vestigial phase is intimately related to the amount of symmetry of the defining point group due to inherently growing fluctuations of the order parameter. This might explain the stability of the observed uniaxial-biaxial phases as compared to the yet to be observed other possible forms of generalized nematic order with higher point group symmetries.
\end{abstract}

\maketitle

\section{Introduction}

``Vestigial" or ``mesophases" of matter are a well established part of the canon of spontaneous symmetry breaking \cite{Friedel22}. It might well happen that due to thermal \cite{DeGennesProst95} (or even quantum \cite{KivelsonFradkinEmery1998}) fluctuations a phase is stabilized at intermediate temperatures (or coupling constant at $T=0$) characterized by a symmetry intermediate between the high temperature isotropic phase and the fully symmetry broken phase at low temperature (small coupling constant).  Iconic examples are liquid crystals \cite{DeGennesProst95}, occurring in between the high temperature liquids and the low temperature crystals, characterized by only the breaking of the rotational symmetry (``nematics"), followed potentially by a partial breaking of translations (``smectic" or ``columnar" phases) before full solidification sets in.  

In the general sense of phases of matter that break the isotropy of Euclidean three dimensional space, crystals are completely classified in terms of space groups. Nematics, on the other hand, are in principle classified in terms of all subgroups of $O(3)$: the family of 3D point groups. There are a total of seven infinite axial families and seven polyhedral groups of such symmetries, exhibiting a very rich subgroup hierarchy. For instance, one can contemplate a descendence like $O(3) \rightarrow SO(3) \rightarrow I \rightarrow T \rightarrow \cdots \rightarrow D_2 \rightarrow C_2 \rightarrow C_1$. Accordingly, in principle it is allowed by symmetry to realize a very rich hierarchy of rotational vestigial phases, where upon lowering temperature phases in this symmetry hierarchy would be realized one after the other. 

In experimental reality this is not encountered \cite{Lehmann1889, DeGennesProst95}. Nearly all of the vast empirical landscape of liquid crystals deals with one particular form of nematic order: the uniaxial nematic characterized by the $D_{\infty h}$ pointgroup with ``rod-like" molecules or mesogens that line up in the nematic phase.  Another well established form is the ``biaxial nematic" formed from platelets with three inequivalent director axes, characterized by the $D_{2h}$ point group symmetry \cite{Freiser70, Straley74, YuSaupe80, MadsenEtAl04, AcharyaPrimakKumar04, MerkelEtAl04, Matteis2008, Tschierske10, BiaxialBook2015}. $D_{2h}$ is a subgroup of $D_{\infty h}$ and it is well understood that the uniaxial nematic can be a vestigial mesophase that can occur in between the isotropic and biaxial phase. In order for such vestigial rotational sequences to occur special microscopic conditions are required: dealing with molecule-like mesogenic constituents, special anisotropic interactions have to be present.

\begin{figure}
\includegraphics[width=0.45\textwidth]{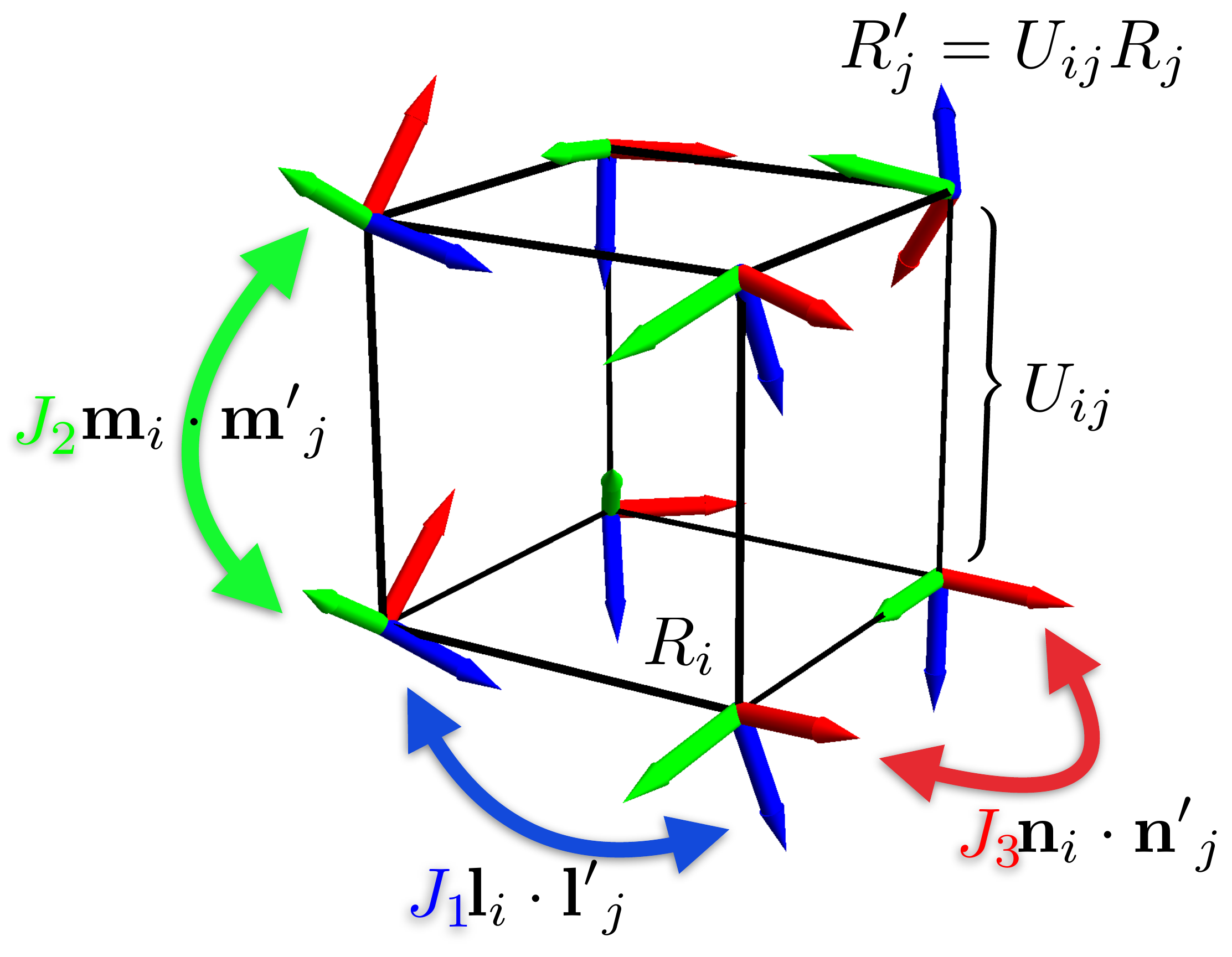}
\caption{Point-group symmetric orientational degrees of freedom $R_i, R_j'=U_{ij}R_j$ on a lattice with local identifications $R_{i,j} \simeq \Lambda_i R_{i,j}$, for $\Lambda_i \in G$, the associated gauge fields $U_{ij} \simeq \Lambda_i U_{ij} \Lambda_j^T$ on links $\corr{ij}$, and the nearest-neighbor Hamiltonian $\mathrm{Tr}~[R^T_i \mathbb{J}U_{ij}R_j]$ between triads $\vek{n}^{\alpha}_i = \{\vek{l}_i, \vek{m}_i, \vek{n}_i\}$, $\vek{n}^{\prime \alpha}_j = \{\vek{l}^{\prime}_j, \vek{m}^{\prime}_j, \vek{n}^{\prime}_j\}$ with parameters $\mathbb{J} = \mathrm{diag}(J_1,J_2,J_3)$. For clarity we show the couplings of the triads on several different nearest neighbors sites $i,j$. For more details on lattice model and the gauge theoretical description of nematics, see Section \ref{sec:lattice model}.}
\label{fig:J1-J2-J3}
\end{figure}

More concretely, in terms \jaakko{of a theory with} lattice regularization, the degrees of freedom of the coarse-grained orientational constituents can be parametrized in terms of an $O(3)$-rotation matrix $R_i = (\vek{l}_i \ \vek{m}_i \ \vek{n}_i)^T$, i.e. an orthonormal triad $\vek{n}^{\alpha}_i = \{\vek{l}_i, \vek{m}_i, \vek{n}_i\}_{\alpha =1,2,3}$ in the body-fixed frame of the mesogen \cite{LiuEtAl2015b}. The orientational interaction between the mesogens is in general determined by their relative orientation of nearest neighbor sites $i,j$ and therefore a function of the relative direction cosines, i.e. $H_{ij} \sim - \mathrm{Tr}~ [R_i^T \mathbb{J} R'_j ]= -\sum_{\alpha \beta} \mathbb{J}^{\alpha\beta} \vek{n}^{\alpha}_i \cdot \vek{n'}^{\beta}_j$, where $\mathbb{J}^{\alpha \beta}$ is a symmetric matrix, see Fig. \ref{fig:J1-J2-J3}. It turns out that without loss of generality this matrix can be diagonalized and the eigenvalues $J_1,J_2,J_3$ of $\mathbb{J}$  characterize the interaction in terms of three perpendicular axes. Furthermore, the local axes $\vek{n}^{\alpha}_k = \{\vek{n}^{\alpha}_{i},\vek{n'}^{\alpha}_j\}_{k\in \corr{ij}}$ are identified under the local point-group symmetries $\Lambda_i \in G$ in their body-fixed frame as $\vek{n}_k^{\alpha} \simeq \Lambda_i^{\alpha\beta}\vek{n}^{\beta}_{k}$ and the form the matrix $\mathbb{J}$ is constrained by the point-group symmetry $G$ of the mesogens, see Section \ref{sec:lattice model}. It is the case that the point groups are classified into two classes: the seven finite polyhedral groups $T, T_h, T_d, O, O_h, I, I_h$ that only allow for an isotropic $\mathbb{J} = J\id$ and the seven infinite families of groups $C_n, C_{nv}, C_{nh}, S_{2n}, D_{n}, D_{nh}, D_{nd}$, where anisotropy in $J_1-J_2-J_3$ should be in general expected since it is allowed by the symmetries. 

Aside from the uniaxial $D_{\infty h}$-nematic with a \emph{single} director axis, the main focus regarding such anisotropies has been on a particular point group symmetry generalizing the uniaxial ordering to three dimensions --- the biaxial $D_{2h}$ ``platelet" with three inequivalent director axes. The expectation is then that the biaxial phase is stabilized by sufficient anisotropy in the constituents and/or interactions \cite{Freiser70, Straley74, SonnetVirgaDurand03, BiaxialBook2015}. 

These point groups have been \jaakko{the main focus of attention in mesogenic systems} and we are only aware of a few other additional point groups that have been considered in similar detail. That is, besides the $D_{2h}$ symmetry, only mesophases of $C_{2v}$ point-group-symmetric ``banana-shaped" constituents have recently been investigated in some detail \cite{Takezoe06, Tschierske10} in experimental systems, subsequently followed by theoretical considerations \cite{LubenskyRadzihovsky2002, Mettout2005,BatesLuckhurst05}, as well as theretical studies of other mesogenic symmetries \cite{Mettout2006, GorkunovEtAl10, LuckhurstEtAl11}. However, in these systems the $C_{2v}$ constituents seem to organize into complicated mesogenic aggregates in the observed liquid crystals, thereby \jaakko{many of the systems form} columnar and smectic phases \cite{Takezoe06}. 

As we will discuss in detail in the next section, the symmetry structure and anisotropic interactions that are behind the $D_{2h}$ uniaxial-biaxial phase descendence are actually perfectly compatible with all axial groups! As a consequence, the generalization of the special uniaxial-biaxial type of vestigial symmetry lowering is possible for this vast number of symmetries. In fact,  the axial groups roughly divide into two subclasses in this particular regard.  $D_{2h}$ belongs to the symmetry class\josko{es that are } characterized by a horizontal mirror plane and the $J_1-J_2-J_3$ type of anisotropy allows for just a single vestigial phase where fluctuations restore rotational symmetries in the mirror plane, which is always $D_{\infty h}$ the uniaxial nematic. However, in the other case, such a mirror plane is lacking and we show in section II that this makes possible a second generic ``biaxial$^*$" phase with an extra mirror symmetry along the main axis compared to the original low temperature biaxial phase. 

Aside from pure symmetry considerations, the next question is how do the stability of the vestigial phase(s) and the fully ordered phase depend on general conditions such as the couplings and the nature of the point group symmetry of the constituents? As we discussed elsewhere in much detail \cite{NissinenEtAl16}, the order parameter theories of `generalized nematics' characterized by symmetries beyond the simple $D_{\infty h}, D_{2h}$ are barely explored. The difficulty is with the complicated tensor structure of these order parameters.  We introduced an extremely convenient mathematical formalism, borrowed from high energy physics, to address these matters: $O(3)$ matrix matter coupled to discrete non-Abelian point group $G$ gauge theory. On the technical side, the gauge-theoretic framework is a convenient device to construct the explicit order parameter tensors \cite{NissinenEtAl16}, but we also found that it is remarkably powerful to address the order-out-of disorder physics behind the occurrence of the vestigial phases \cite{LiuEtAl2015b}. We found thermal fluctuations of unprecendented strength lowering the transition temperatures to very low values in case of the most symmetric point groups ($T, O, I$), giving rise to a natural occurrence of a spontaneous vestigial chiral phase dealing with chiral point groups. How does this motive relate to the present context of  ``generalized" uniaxial-biaxial sequences? 

It is actually the case that the $J_1 - J_2 - J_3$ type of anisotropy that arises in the gauge theory \jaakko{allows one to incorporate the generalized biaxial transitions in a natural manner}, thereby making it possible to study such transitions with remarkable ease. We will discuss this in more detail in Section III how to use the gauge theory to compute quantitative phase diagrams. \josko{As expected, } we recover the generic topology of the phase diagrams as \josko{a} function of the anisotropy parameters. \jaakko{The advantage is} that in the gauge theory one can compare apples with apples and pears with pears in the sense that the strength of the microscopic interactions including their anisotropy can be kept the same, facilitating a qualitative comparison of the phase diagrams for different point groups symmetries.  {\em The conclusion is that the stability region of the vestigial uniaxial phase grows rapidly as a function of increasing symmetry of the point group, \josko{which surpresses} the fully ordered generalized biaxial phases considerably.} 

This mirrors the general motive that we already identified in the context of the chiral vestigal phases \cite{LiuEtAl2015b}: for the more symmetric point groups the thermal fluctuations grow in severity. This has on the one hand the effect of suppressing the ordering temperature of the fully ordered generalized biaxial phases, while at the same time the vestigial phase to a degree {\em profits} from the thermal fluctuations. As we will further discuss in the conclusion section, this raises the question whether for systems made from constituents characterized by highly symmetric point-groups it will be ever possible to find the fully ordered phases before other mesophases and/or solidification sets in (these are beyond the description of our orientational lattice model). Any microscopic anisotropy might well render the vestigial uniaxial phase to be only one that can be realized.

The remainder of this paper is organized as follows. In Section~\ref{sec:symmetries} we discuss the possible axial nematic phase transitions in terms of symmetries. For a realization of these phase transitions, we review the lattice gauge theory model and define its anisotropic coupling parameters in Section~\ref{sec:lattice model}. Section~\ref{sec:MC_results} is devoted for the phase diagrams and phase transitions obtained in Monte Carlo simulations. We conclude with an outlook in Sec.~\ref{sec:conclusions}.

\section{The structure of nematic order parameters and generalized biaxial transitions} \label{sec:symmetries}

Three dimensional generalized nematics break the rotational group $O(3)$ down to a three-dimensional point group.
By the Landau-de Gennes symmetry paradigm, phase transitions between any two nematic phases related by the subgroup structure of $O(3)$ are allowed, in addition to the transitions between the isotropic $O(3)$ phase and a generalized nematic phase. In this section, we will show that the order parameter structure of axial nematics provides a natural way to realize a some of the symmetry allowed transitions. In Section \ref{sec:lattice model} we then discuss how to realize these phase transitions by tuning the couplings in our gauge theoretical setup \cite{LiuEtAl2015b}.

\subsection{Point groups and nematic order parameters} \label{sec:J_op_structure}

Three-dimensional point groups are classified as seven finite polyhedral groups, $\{T, T_d, T_h, O, O_h, I, I_h  \}$, and seven infinite families of axial groups, $\{C_n, C_{nv}, S_{2n}, C_{nh}, D_n, D_{nh}, D_{nd} \}$ \cite{SternbergBook, Michel2001}.
The associated nematic order parameters are tensors that are invariant under the given point-group symmetry.
A full classification of these order parameters and their derivation is given in our recent paper \cite{NissinenEtAl16}. For the present purposes we therefore review the results that are of importance in the following.

Three dimensional orientation can be parametrized in terms of a $O(3)$ matrix
\begin{align} \label{eq:J_R}
R = \big( \mathbf{l} \quad \mathbf{m} \quad \mathbf{n} \big)^T.
\end{align}
The rows $\mathbf{n}^{\alpha} = \{ \mathbf{l}, \mathbf{m}, \mathbf{n} \}$ of $R$ form an orthonormal triad and satisfy the additional $O(3)$ constraint
\begin{align} 
\sigma = \det R = \epsilon_{abc}(\mathbf{l}\otimes \mathbf{m} \otimes \mathbf{n})_{abc} =  \mathbf{l} \cdot ( \mathbf{m} \times \mathbf{n} ) = \pm 1, \label{eq:sigma}
\end{align}
where $\sigma$ is the chirality or handedness of the triad $\vek{n}^{\alpha}$ associated with $R$.

The order parameter tensors are constructed from tensor products of $R$ and we use the point group conventions of Ref. \onlinecite{NissinenEtAl16}. In case of the polyhedral nematics $G=\{T, T_d, T_h, O, O_h, I, I_h  \}$, the general form of the order parameter is given by
$\mathbb{O}^G = \{ \mathbb{O}^{G}[\mathbf{l} \ \mathbf{m} \ \mathbf{n}], \sigma \}$, where $\mathbb{O}^{G}[\mathbf{l} \ \mathbf{m} \ \mathbf{n}]$ describes the orientational order of the phase and $\sigma$ is a chiral order parameter needed for the proper polyhedral groups $\{T, O, I \}$. The polyhedral groups have several higher order rotation axes and transform the triads $\{ \mathbf{l}, \mathbf{m}, \mathbf{n} \}$ irreducibly, and in these cases we only need one  tensor to describe the orientational order \cite{NissinenEtAl16}.

On the other hand, the axial groups $\{C_n, C_{nv}, S_{2n}, C_{nh}, D_n, D_{nh}, D_{nd} \}$ are defined with respect to a symmetry plane involving rotations and/or reflections and a perpendicular, axial direction. \jaakko{Their irreducible representations are in general one- \josko{or} two-dimensional}. Correspondingly, the order parameter tensors of  the axial point groups have the general structure $ \mathbb{O}^G = \{  \mathbb{A}^G, \mathbb{B}^G, \sigma \}$, where $\mathbb{A}^G$ defines the ordering related to the orientation of the primary axial axis perpendicular to the symmetry plane and
$\mathbb{B}^G$ describes the in-plane ordering. We refer to $\mathbb{A}$ as the axial order and $\mathbb{B}$ as the in-plane (or just biaxial) order \cite{NissinenEtAl16}. Similarly, $\sigma$ is the chiral ordering for the proper axial groups $\{C_{n}, D_n \}$. Note that the $O(3)$ constraints can reduce the number of independent order parameter tensors in the set $\{\mathbb{A}^G, \mathbb{B}^G, \sigma\}$ \cite{NissinenEtAl16}.
Following the conventions in Ref. \onlinecite{NissinenEtAl16}, $\mathbf{n}$ is chosen always to be along the primary, \jaakko{axial} axis.  \jaakko{It follows that} the axial order parameter tensor $\mathbb{A}^G = \mathbb{A}^G [\mathbf{n}]$ depends only on $\mathbf{n}$ by construction. Similarly, the in-plane order parameter $\mathbb{B}^G = \mathbb{B}^G [\mathbf{l}, \mathbf{m}]$ depends only on $\{ \mathbf{l}, \mathbf{m} \}$ for the symmetries $G = \{C_n, C_{nv}, C_{nh}, D_n, D_{nh} \}$, but is a  tensor polynomial $\mathbb{B}^G = \mathbb{B}^G [\mathbf{l}, \mathbf{m}, \mathbf{n}]$ of all the three triads for the symmetries $\{ S_{2n}, D_{nd}\}$ \jaakko{with rotoreflections}. We have discussed these ordering tensors in Ref. \cite{NissinenEtAl16}, but for the convenience of the readers, we show the relevant selection of \jaakko{order parameter tensors for the axial groups} in Table \ref{table:J_ops}.

Moreover, because of the common structure of the axial point groups, the tensors $\mathbb{A}^G$ and $\mathbb{B}^G$  are not unique to a given symmetry, though the axial point group ordering can be uniquely defined by the full set of order parameters $\{  \mathbb{A}^G, \mathbb{B}^G, \sigma \}$. For instance, the symmetry groups $C_n$ and $C_{nv}$ do not transform the primary axis $\mathbf{n}$, thus the axial ordering tensor for symmetries in these types is simply a vector,
\begin{align} \label{eq:J_Cinfv_op}
\mathbb{A}^{C_n}[\mathbf{n}] = \mathbb{A}^{C_{nv}}[\mathbf{n}] = \mathbb{A}^{C_{\infty v}}[\mathbf{n}] = \mathbf{n},
\end{align}
where $C_{\infty} \cong SO(2)$ is the continuous limit of $C_n$ and $C_{\infty v} \cong O(2)$ is the continuous limit of $C_{nv}$.
The symmetries $\{ S_{2n}, C_{nh}, D_n, D_{nh}, D_{nd} \}$, however, transform $\mathbf{n}$ to $-\mathbf{n}$, and therefore have the same axial ordering tensor
\begin{align}\label{eq:J_Dinfh_op}
\mathbb{A}^{D_{\infty h}}[\mathbf{n}] & = \mathbb{A}^{C_{\infty h}}[\vek{n}] = \mathbb{A}^{C_{nh}}[\mathbf{n}] = 
\mathbb{A}^{D_{n}}[\mathbf{n}] = \mathbb{A}^{D_{nh}}[\mathbf{n}] 
\nonumber \\ 
& =  \mathbb{A}^{D_{nd}}[\mathbf{n}]  = \mathbb{A}^{S_{2n}}[\mathbf{n}] = \mathbf{n} \otimes \mathbf{n} -\frac{1}{3}\id,
\end{align}
which is just the well-known director order parameter of $D_{\infty h}$-uniaxial nematics. Note that $D_{\infty h}$ can be considered as the continuous limit of the finite groups $D_{nh}$, and $D_{nd}$, whereas $C_{\infty h}$ arises from the limit of $C_{nh}$ and $S_{2n}$.
Similarly, axial nematics with the same $n$-fold in-plane symmetries have the same ordering tensor $\mathbb{B}$,
\begin{align}\label{eq:J_b_op}
& \mathbb{B}^{C_n}[\mathbf{l}, \mathbf{m}] = \mathbb{B}^{C_{nh}}[\mathbf{l}, \mathbf{m}],
\nonumber \\ 
& \mathbb{B}^{C_{nv}}[\mathbf{l}, \mathbf{m}] =  \mathbb{B}^{D_{n}}[\mathbf{l}, \mathbf{m}]  = \mathbb{B}^{D_{nh}}[\mathbf{l}, \mathbf{m}].
\end{align}
Note that, though the axial and the biaxial ordering tensors are distinct and transform irreducibly, they are not completely independent due to the $O(3)$ constraints of orthonormality and Eq. \eqref{eq:sigma}.

\setlength{\tabcolsep}{6pt}
\renewcommand{\arraystretch}{2}
\begin{table*}[!htb]
\centering
\caption{{\bf Generalized biaxial phase transitions.} The first column specifies the generalized nematic symmetries and the second column the minimal set of order parameter tensors for their characterization. 
Relations of the order parameters given by Eqs. \eqref{eq:J_Cinfv_op}--\eqref{eq:J_b_op} are indicated. For the explicit form of these order parameters see Ref. \cite{NissinenEtAl16}.
The third and fourth column show the order parameter tensors involved in the generalized biaxial-uniaxial transitions in Eq. \eqref{eq:aixal_transition} and the biaxial-biaxial$^*$ transitions in Eq. \eqref{eq:bb_transition}, respectively. The symbol ``$\rightarrow$'' indicates the replacement of an order parameter that becomes non-vanishing for the higher symmetry biaxial* (or uniaxial*) phases.} \label{table:biaxial_transitions}
\begin{tabular}{ | c | c | c |  c |  }
    \hline
    \hline
   \parbox{1.5cm}{\centering{\bf Symmetry }}
	& \parbox{5.5 cm}{\centering{\bf Order  Parameters }}   
   & \parbox{3.5 cm}{\centering{\bf Uniaxial-biaxial \\ Transitions}}
   & \parbox{3.5 cm}{\centering{\bf Biaxial-biaxial* (Uniaxial-uniaxial*)  Transitions}}  \\ \hline
   
   $C_n$
   & $ \mathbb{A}^{C_{n}} = \mathbb{A}^{C_{\infty v}}[\mathbf{n}]$, 
   		$ \mathbb{B}^{C_{n}} = \mathbb{B}^{C_{nh}}[\mathbf{l},\mathbf{m}] $, $\sigma$
   &  $\mathbb{B}^{C_{nh}}[\mathbf{l},\mathbf{m}]$, $\sigma$
   &  $\mathbb{A}^{C_{\infty v}}[\mathbf{n}] \rightarrow \mathbb{A}^{D_{\infty h}}[\mathbf{n}]$, $\sigma$
   \\ \hline
   
   $C_{nv}$
   & $ \mathbb{A}^{C_{nv}} = \mathbb{A}^{C_{\infty v}}[\mathbf{n}] $, 
   		$\mathbb{B}^{C_{nv}} = \mathbb{B}^{D_{nh}}[\mathbf{l},\mathbf{m}] $
   & $\mathbb{B}^{D_{nh}}[\mathbf{l},\mathbf{m}]$ 
   &  $\mathbb{A}^{C_{\infty v}}[\mathbf{n}] \rightarrow \mathbb{A}^{D_{\infty h}}[\mathbf{n}]$
   \\ \hline
   
   $S_{2n}$
   & $ \mathbb{A}^{S_{2n}} = \mathbb{A}^{D_{\infty h}}[\mathbf{n}] $, 
   		$\mathbb{B}^{S_{2n}}[\mathbf{l},\mathbf{m},\mathbf{n}] $
   & $\mathbb{B}^{S_{2n}}[\mathbf{l},\mathbf{m},\mathbf{n}] $
   &  $\mathbb{B}^{S_{2n}}[\mathbf{l},\mathbf{m},\mathbf{n}]  \rightarrow \mathbb{B}^{C_{2nh}}[\mathbf{l},\mathbf{m}]$
   \\ \hline
   
   $C_{nh}$
   & $ \mathbb{A}^{C_{nh}} = \mathbb{A}^{D_{\infty h}}[\mathbf{n}] $, 
   		$\mathbb{B}^{C_{nh}}[\mathbf{l},\mathbf{m}] $
   & $\mathbb{B}^{C_{nh}}[\mathbf{l},\mathbf{m}] $
   &  {\rm None}
   \\ \hline
   
   $D_{n}$
   & $ \mathbb{A}^{D_{n}} = \mathbb{A}^{D_{\infty h}}[\mathbf{n}] $, 
   		$\mathbb{B}^{D_{n}} =\mathbb{B}^{D_{nh}}[\mathbf{l},\mathbf{m}] $, $\sigma$
   & $\mathbb{B}^{D_{nh}}[\mathbf{l},\mathbf{m}] $, $\sigma$
   &   $\sigma$
   \\ \hline

 $D_{nh}$
   & $ \mathbb{A}^{D_{nh}} =\mathbb{A}^{D_{\infty h}}[\mathbf{n}] $, 
   		$\mathbb{B}^{D_{nh}}[\mathbf{l},\mathbf{m}] $
   & $\mathbb{B}^{D_{nh}}[\mathbf{l},\mathbf{m}] $
   &   {\rm None}
   \\ \hline

 $D_{nd}$
   & $ \mathbb{A}^{D_{nd}} = \mathbb{A}^{D_{\infty h}}[\mathbf{n}] $, 
   		$\mathbb{B}^{D_{nd}}[\mathbf{l},\mathbf{m},\mathbf{n}] $
   & $\mathbb{B}^{D_{nd}}[\mathbf{l},\mathbf{m},\mathbf{n}] $
   &   $\mathbb{B}^{D_{nd}}[\mathbf{l},\mathbf{m},\mathbf{n}] \rightarrow \mathbb{B}^{D_{2nh}}[\mathbf{l},\mathbf{m}]$
   \\ \hline
   
  $C_{\infty v}$
   & $ \mathbb{A}^{C_{\infty v}}[\mathbf{n}]$
   & {\rm None}
   &  $\mathbb{A}^{C_{\infty v}}[\mathbf{n}] \rightarrow \mathbb{A}^{D_{\infty h}}[\mathbf{n}]$
   \\ \hline

     $D_{\infty h}$
   & $ \mathbb{A}^{D_{\infty h}}[\mathbf{n}] $
   & {\rm None}
   &  {\rm None}
   \\ \hline  
   
   \hline 
\end{tabular}

\end{table*}

\subsection{Generalized biaxial phases and transitions}\label{sec:biaxial_transition}

The distinction between the primary axis $\mathbf{n}$ and the in-plane axes $\mathbf{l}$ and $\mathbf{m}$ for axial nematics allows the disordering of the axial and in-plane order separately.  

A familiar example is the biaxial-uniaxial-isotropic liquid transitions of $D_{2h}$-biaxial liquid crystals \cite{Freiser70, Alben73, Straley74, BiscariniZannoni1995, AllenderLeeHafiz85, AllenderLonga08}. The order parameter tensors of the $D_{2h}$ nematic are defined by two linearly independent rank-$2$ tensors, 
$\mathbb{O}^{D_{2h}} = \{\mathbb{A}^{D_{2h}}[\mathbf{n}], \mathbb{B}^{D_{2h}}[\mathbf{l},\mathbf{m}] \}$,
where $\mathbb{A}^{D_{2h}}[\mathbf{n}] $ has been given in Eq. \eqref{eq:J_Dinfh_op},
 and  $\mathbb{B}^{D_{2h}}[\mathbf{l},\mathbf{m}]$ is the well-known biaxial order parameter,
\begin{align}
\mathbb{B}^{D_{2h}} [\mathbf{l},\mathbf{m}] = \mathbf{l} \otimes \mathbf{l} - \mathbf{m} \otimes \mathbf{m}.
\end{align}
In terms of the symmetries, the biaxial nematic order allows for the phase transitions
\begin{align} 
D_{2h} \rightarrow D_{\infty h} \rightarrow O(3),
\end{align}
with the uniaxial phase occurring before the isotropic liquid. That is, upon increasing temperature, the biaxial order is destroyed first leading to the restoration of the in-plane $O(2)$ symmetry of uniaxial nematics before the transition to the fully disordered isotropic phase takes place.

Given the general order parameter structure of axial nematics discussed in Section \ref{sec:J_op_structure}, this transition sequence can be directly generalized to other axial symmetries. We will refer to the associated phase transitions as \emph{generalized biaxial transitions}. By first destroying the in-plane order $\mathbb{B}$, the following generalized biaxial-uniaxial transition can be induced
\begin{align} \label{eq:aixal_transition}
 & C_n, C_{nv} \rightarrow C_{\infty v} , \nonumber \\
 & S_{2n}, C_{nh}, D_{n}, D_{nh}, D_{nd} \rightarrow D_{\infty h}.
\end{align}
Note that in these transitions we consider situations where the in-plane order has been completely disordered, leading to full $O(2)$ symmetry. Thus the chiral order $\sigma$ for proper groups $C_n$ and $D_n$ has been simultaneously lost. Nevertheless, we can in principle also have the restorations of only the in-plane $SO(2)$ symmetry with the transitions 
\begin{align}
 C_n \rightarrow C_{\infty}, \ 
 D_{n} \rightarrow D_{\infty}.
\end{align}
where the chirality $\sigma$ does not disorder \cite{LiuEtAl2015b}. However, since $\sigma$ is a composite order parameter of $\{ \mathbf{l}, \mathbf{m}, \mathbf{n}\}$ featuring also some in-plane ordering, these transitions require more fine tuning in comparison to those in Eq. \eqref{eq:aixal_transition}.

In the opposite limit, if the in-plane order with order parameter $\mathbb{B}$ is sufficiently strong in comparison to the axial ordering $\mathbb{A}[\vek{n}]$, we can disorder the primary axis $\mathbf{n}$ without destroying the in-plane order upon increasing the temperature. Note that due to the $O(3)$ constraints on the triads, the axial ordering is never fully independent in the presence of the perpendicular in-plane ordering that fixes $\vek{n}$ up to sign. Therefore, upon disordering the axial order, the symmetry of the phase is augmented by
\begin{align}
\sigma_h = \left(
\begin{array}{ccc}
 1 & 0 & 0 \\
  0& 1 & 0 \\
 0 & 0 & -1
\end{array}
\right),
\end{align}
which is a simply a reflection with respect to the $(\mathbf{l}, \mathbf{m})$ plane that acts trivially on the in-plane ordering. Other symmetry operations transforming $\mathbf{n}$ to $-\mathbf{n}$, such as the inversion or a two-fold rotations about an axis in the $(\mathbf{l},\mathbf{m})$-plane, however, will simultaneously transform the in-plane order. If such symmetries belong to the original symmetry group $G$, they will lead to enhanced in-plane symmetries in combination with $\sigma_h$. Therefore the new symmetries due to the disordering of the axial order $\mathbb{A}^G[\vek{n}]$ are generated by the elements $G^* = \corr{G, \sigma_h}$, leading schematically to \emph{either} the direct product structure $G^*= G'  \times \{\id, \sigma_h\}$ \emph{or} the semi-direct structure $G^* = G' \ltimes \{\id, \sigma_h \}$, where $G'$ can be an $n$-fold or $2n$-fold rotational group. These are transitions between phases with different ``biaxial" orders $\mathbb{B}^G$ and $\mathbb{B}^{G^*}$ will be for convenience referred to as biaxial-biaxial$^*$ transitions, where the subscript in $G^*$ denotes the presence of the additional reflections in comparison with the low temperature symmetries $G$. The behavior of the associated orders in the generalized uniaxial-biaxial transitions Eq. \eqref{eq:aixal_transition} and biaxial-biaxial$^*$ transition are summarized in Table \ref{table:biaxial_transitions}.

More specifically, in the``biaxial-biaxial$^*$" \jaakko{phase transition} the disordering of the primary axis with order parameter $\mathbb{A}^G[\vek{n}]$ will lead to the phase transition of the generalized nematics with symmetries $\{C_n, C_{nv}, S_{2n}, D_n, D_{nh}, D_{nd} \}$ 
\begin{align} \label{eq:bb_transition}
& C_n \rightarrow C_{nh}, \nonumber \\
& S_{2n} \rightarrow C_{2n h}, \nonumber \\
&  C_{nv}, D_{n} \rightarrow D_{nh}, \nonumber \\
& D_{nd} \rightarrow D_{2nh},
\end{align}
as follows from the subgroup structure of $O(3)$. \jaakko{Since $\sigma_h$ is already contained in the groups $C_{nh}$ and $D_{nh}$, the biaxial* phase is not present for these nematics}.

Indeed, we see that these transitions have more interesting features than the generalized uniaxial-biaxial transitions in Eq. \eqref{eq:aixal_transition}, because $\sigma_h$ may be ``fused'' to the parent symmetries via a direct product or semi-direct product, leading to different effects on the original order.
For instance,  for $C_n$ and $C_{nv}$ nematics, whose axial order parameter $\mathbb{A}^G[\vek{n}]$ is simply  the vector $\mathbf{n}$, disordering the primary axis in the presence of the in-plane order, i.e. adding the extra symmetry generator $\sigma_h$, will simply lift \jaakko{the vector order parameter} to a director. Consequently, the original axial order is destroyed, but a new axial order will persist as long as $\mathbb{B}$ is ordered and subsequently leads to the nematic order $\mathbb{B}^{G^*}$.

Moreover, for $D_n$ nematics the axial order is already fixed by the in-plane $\mathbb{B}$ with $C_n$ rotations up to a sign, as well as being invariant under the dihedral $\pi$-rotations $\vek{m} \to -\vek{m}, \mathbf{n} \rightarrow -\mathbf{n}$. Therefore, upon increasing the temperature and disordering the primary axis, i.e. adding $\sigma_h$ to the symmetries of the phase, the transition $D_{n} \to D_{nh}$ occurs, ensuring the vanishing of the chiral order parameter $\sigma$. This is accompanied, perhaps counter intuitively, by the axial order parameter $\mathbb{A}[\vek{n}]$ still being non-zero, albeit with reduction in its magnitude due to the higher temperature.

Last but not the least, in the cases of $S_{2n}$ and $D_{nd}$ nematics with rotoreflection symmetries, disordering $\vek{n}$ and promoting $\sigma_h$ to the axial axis lifts their in-plane structure to a higher in-plane symmetry, since the biaxial order parameter for these symmetries is a function of all the three triads, $\mathbb{B}^{S_{2n},D_{nd}} = \mathbb{B}^{S_{2n},D_{nd}}[\mathbf{l}, \mathbf{m}, \mathbf{n}]$.

\section{Lattice realization of generalized biaxial transitions} \label{sec:lattice model}

The generalized biaxial transitions in Eq. \eqref{eq:aixal_transition} and Eq. \eqref{eq:bb_transition} generalize the biaxial-uniaxial transition of $D_{2h}$ nematics into a much broader class.
These transitions can be readily addressed using the gauge-theoretical description for generalized nematics as introduced in Ref. \cite{LiuEtAl2015b}. We now recollect the model, to subsequently show how anisotropic couplings that do not break any symmetries serve as tuning parameters for the generalized unaxial-biaxial phase transitions in Sec. \ref{sec:biaxial_transition}.

\subsection{Gauge theoretical description of  generalized nematics} \label{sec:J_gauge_model}

In Ref. \cite{LiuEtAl2015b}, we introduced a gauge theoretical setup to describe generalized nematic order with arbitrary three-dimensional point group symmetry. \jaakko{In the gauge theoretical approach, instead of directly dealing with order parameter tensors, the symmetry of three dimensional nematic orders is realized by a point-group-symmetric gauge theory coupled to $O(3)$ matter}. The model is in general a discrete non-Abelian lattice gauge theory with \jaakko{$O(3)$-matter in the fundamental representation}, generalizing the $\integers_2$ Abelian Lammert-Rokshar-Toner gauge theory for the uniaxial $D_{\infty h}$-nematic \cite{LammertRoksharToner93, LammertRoksharToner95}. The nematic phase and the isotropic phase are realized by the Higgs phase and the confined phase of the gauge theory, respectively.

The model is defined by the Hamiltonian \cite{LiuEtAl2015b},
\begin{align}
H & = H_{\rm Higgs} + H_{\rm gauge}, \label{eq:J_gauge_theory}
\\
H_{\rm Higgs} & = -\sum_{\langle ij \rangle} \textrm{Tr}~\big[R_i^T \mathbb{J} U_{ij} R_j\big], \label{eq:J_Higgs}
\\
H_{\rm gauge} & = - \sum_{\square} \sum_{\mathcal{C}_\mu} K_{\mathcal{C}_\mu} \delta_{\mathcal{C}_{\mu}}(U_{\square}) \mathrm{Tr}\big[U_{\square}\big].
\end{align}
The matter fields $\{R_i\}$ live on the sites of a cubic lattice and are $O(3)$ matrices, as in Eq. \eqref{eq:J_R}. 
The gauge fields $\{U_{ij}\}$ are elements of the point group $G$ and live on the links $\langle ij \rangle$. In the Hamiltonian, $H_{\rm Higgs}$ is a Higgs term \cite{FradkinShenker79} describing interactions between the matter fields $R_i$ and gauge fields $U_{ij}$, parametrized by the coupling matrix $\mathbb{J}$ determining how the local axes $\{ \mathbf{n}^{\alpha}_i \}$ are coupled, see Fig. \ref{fig:J1-J2-J3}.
The Hamiltonian in Eq. \eqref{eq:J_gauge_theory} is invariant under local gauge transformations
\begin{align}
R_i \to \Lambda_i R_i, \quad U_{ij} \to \Lambda_i U_{ij} \Lambda_j^T, \quad \forall \Lambda_i \in G,
\end{align}
which leads to the identifications 
\begin{align}
R_i \simeq \Lambda_i R_i, \quad \vek{n}_i^\alpha \simeq \Lambda_i^{\alpha\beta}\vek{n}_i^\beta, \quad \Lambda_i\in G. \label{eq:mesogens}
\end{align}
Thus  $H_{\rm Higgs}$ effectively models the orientational interaction between two $G$-symmetric ``mesogens'' \cite{LiuEtAl2015b}. In addition, $H_{\rm Higgs}$ has the global $O(3)$-rotation symmetry 
\begin{align}
R_i \rightarrow R_i \Omega^T, \ \Omega \in O(3). \label{eq:global symmetry}
\end{align}
Since gauge symmetries cannot be broken \cite{Elitzur75}, the fully ordered Higgs phase of $H_{\rm Higgs}$ will develop long range order characterized by $G$-invariant tensor order parameters and thus realizes spontaneous symmetry breaking of Eq. \eqref{eq:global symmetry} from an isotropic $O(3)$ liquid phase to \jaakko{a generalized nematic phase} \cite{LiuEtAl2015b, NissinenEtAl16}.

The term $H_{\rm gauge}$ in the Hamiltonian describes a point-group-symmetric gauge theory \cite{Kogut79}. The term
$U_{\square} = \prod_{\corr{ij} \in \square} U_{ij}$ denotes the oriented product of gauge fields around a plaquette $\square$ and represent the local gauge field configuration on the lattice.
 Plaquettes with non-trivial flux $U_{\square} \neq \id$ represent non-vanishing gauge field strength.
 Due to the gauge symmetries, gauge fluxes in the same conjugacy class are physically equivalent, therefore the coupling $K_{\mathcal{C}_{\mu}}$ is a function on the conjugacy classes $\mathcal{C}_{\mu}$ of the group $G$.
These gauge fluxes are elements of the point group $G$ and correspond to the Volterra defects in nematics \cite{LiuEtAl2015b, KlemanFriedel08}, and thus $K_{\mathcal{C}_{\mu}}$ equivalently assigns a finite core energy to the topological defects in the nematic \cite{LammertRoksharToner95}.
However, for the purpose of realizing the generalized biaxial transitions in Eqs. \eqref{eq:aixal_transition} and \eqref{eq:bb_transition}, the Hamiltonian $H_{\rm Higgs}$ is sufficient and for simplicity we will take $K_{\mathcal{C}_{\mu}} = 0 $ in the following.
  
\subsection{Anisotropic couplings and  generalized biaxial transitions} \label{sec:J_anisotropy}

\setlength{\tabcolsep}{2pt}
\renewcommand{\arraystretch}{1.8}
\begin{table}
\centering
\caption{{\bf Invariant Higgs couplings for point group symmetries.} The nearest-neighbor Higgs coupling $\mathbb{J}$ needs to be invariant under a given three-dimensional point group gauge symmetry $G$, $\Lambda \mathbb{J} \Lambda^T = \mathbb{J}$, $\forall \Lambda \in G$. The possible bilinear forms $\mathbb{J}$ for each symmetry class can be found, e.g. from Ref. \cite{Nye1985}.
} \label{table:J}
\begin{tabular}{ | c | c |  }
    \hline
    \hline
   \parbox{3.5 cm}{\centering{\bf Symmetry  Groups}} 
	& \parbox{4 cm}{\centering{\bf Coupling Matrix }}   
   \\ \hline

   \parbox{3.5 cm} {\centering{$C_1$, $C_i \cong S_2$}}
    & 
		$\left(
		\begin{array}{ccc}
 		J_1 & J_{12} & J_{13} \\
 		J_{12} & J_2 & J_{23} \\
		 J_{13} & J_{23} & J_3
		\end{array}
		\right)$	
   \\ \hline

   \parbox{3.5 cm} {\centering{$C_s \cong C_{1h} \cong C_{1v}$, \\ $C_2, C_{2h}$}}
    & 
		$\left(
		\begin{array}{ccc}
 		J_1 & \ & J_{13} \\
 		\ & J_2 & \ \\
		 J_{13} & \ & J_3
		\end{array}
		\right)$		
   \\ \hline

   \parbox{3.5 cm} {\centering{$C_{2v}, D_2, D_{2h}$}}
    & 
		$\left(
		\begin{array}{ccc}
 		J_1 & \ & \ \\
 		\ & J_2 & \ \\
		 \ & \ & J_3
		\end{array}
		\right)$		
   \\ \hline

  \parbox{3.5 cm} {\centering{$C_{n \geq 3}, C_{(n \geq 3)v}$, \\ 
  								$S_{2(n\geq 2)}$, $C_{(n \geq 3)h}$, \\
  								 $D_{n \geq 3}$, $D_{(n \geq 3)h}$,  $D_{(n \geq 2)d}$}}
    & 
		$\left(
		\begin{array}{ccc}
 		J_1 & \ & \ \\
 		\ & J_1 & \ \\
		 \ & \ & J_3
		\end{array}
		\right)$		
   \\ \hline
  
     \parbox{3.5 cm} {\centering{$T, T_d, T_h$, \\ $O, O_h, I, I_h$}}
    & 
		$\left(
		\begin{array}{ccc}
 		J & \ & \ \\
 		\ & J & \ \\
		 \ & \ & J
		\end{array}
		\right)$		
   \\ \hline

   \hline 
\end{tabular}

\end{table}

In order to analyze the Higgs interaction in terms of the nearest-neighbor local triads $\mathbf{n}_i^{\alpha} =  \{ \mathbf{l}_i, \mathbf{m}_i, \mathbf{n}_i \}$ and $\vek{n}^{\alpha}_j$ identified under \eqref{eq:mesogens}, we can define a local triad vector 
$\mathbf{n}^{\prime \beta}_j = U_{ij}^{\beta \gamma} \mathbf{n}^{\gamma}_j$ at a site $j$, which has been brought (``parallel transported'') into the same local gauge as $\mathbf{n}^{\alpha}_i$ at the site $i$, see Fig. \ref{fig:J1-J2-J3}. In the gauge theory Eq. \eqref{eq:J_Higgs}, each triad $\mathbf{n}^{\alpha}$ represents a local frame of the mesogens and the gauge fields $U_{ij}$ (elements of the point group) on the links encode the relative orientations of the local frames that are ambiguous up to the point-group symmetry of the mesogens. Therefore, in order to analyze the physical orientational interaction between the triads $\mathbf{n}^{\alpha}_i$ and $\mathbf{n}^{\beta}_j$, we need to consider $\mathbf{n}^{\alpha} \cdot \mathbf{n}^{\prime \beta}_j$ that correctly measures the relative orientation. This is mathematically known as the ``parallel transport" of the triad in the gauge potential \cite{Kogut79, FrankelBook} and is hardwired in the gauge theory. The Higgs interaction $H_{\rm Higgs}$ becomes
\begin{align}
H_{\rm Higgs} &= -\sum_{\corr{ij}} \vek{n}^{\alpha}_i \cdot \mathbb{J}^{\alpha \beta} (U_{ij})^{\beta \gamma} \vek{n}^{\gamma}_j
\nonumber \\
& = -\sum_{\corr{ij}} \mathbb{J}^{\alpha \beta} \mathbf{n}^{\alpha}_i \cdot \mathbf{n}^{\prime \beta}_j.
\end{align}
This shows explicitly that the symmetric matrix $\mathbb{J}^{\alpha \beta}$ parametrizes the interaction between the local triads, see Fig. \ref{fig:J1-J2-J3}. Naturally the interaction specified by the bilinear form $\mathbb{J}$ has to respect the symmetry of the underlying ``mesogens'' in Eq. \eqref{eq:mesogens} (i.e. the matter fields in the language of the gauge theory), and needs to satisfy the constraint
\begin{align}
\Lambda \mathbb{J} \Lambda^T  = \mathbb{J}, \ \forall \Lambda \in G
\end{align}
for a given gauge group $G$. This heavily restricts the possible forms of $\mathbb{J}$ that can be found from standard references for crystal symmetry classes (e.g. Ref. \onlinecite{Nye1985}), and we tabulate the results in Table \ref{table:J} for the reader's convenience.

Table \ref{table:J} shows that anisotropic couplings are allowed for axial nematics. This anisotropy is hardwired in the gauge theory Eq. \eqref{eq:J_gauge_theory} and does not break any additional symmetries. Although we have fixed the \emph{local} point group action, i.e. the gauge symmetries, in terms of the triads $\{\vek{l}_i,\vek{m}_i,\vek{n}_i\}$, we can always diagonalize the symmetric matrix $\mathbb{J}^{\alpha \beta}$ by a global redefinition $R_i \to D R_i$, $U_{ij} \to D U_{ij} D^T$. Inspecting the allowed matrices $\mathbb{J}$, the only non-trivial cases are the simple monoclinic symmetries ($C_s, C_2, C_{2h}$), since in the case of $C_1$ and $C_i \simeq S_2 =\{\id, -\id\}$, there are no rotational gauge symmetries $U_{ij}$ to begin with. It is easy to see that the monoclinic symmetries only introduce a common $\pm$ sign in the $\left(\vek{l}, \vek{m}\right)$-plane with the non-diagonal couplings. Therefore without loss of generality we can diagonalize the couplings,
\begin{align}
\mathbb{J} = \left( \begin{matrix} J_1 & & \\ & J_2 & \\ & & J_3 \end{matrix} \right)
\end{align}
with $J_1, J_2, J_3 \geq 0$ for nematic alignment. For the monoclinic symmetries, this requires $J_{13} \leq \sqrt{J_1J_3}$ and we do not consider negative or ``antinematic" couplings \cite{RomanoDeMatteis11, BisiDeMatteisRomano13}.  We further note that the couplings also respect the symmetries of the auxiliary cubic lattice and favor aligment of the triads, leading to homogenous nematic states without any modulation or sublattice structure in the order parameters. Concerning the strength of alignment of the three perpendicular axes, the line of thought can actually be reversed in the sense that we can take couplings $J_1,J_2, J_3$ to be a measure of the effective three dimensionality of the ``mesogens" $R_i$. One realizes that they provide tuning parameters for the phase transitions involving the axial and in-plane ordering.

For the purpose of realizing the transitions in Eq. \eqref{eq:aixal_transition} and Eq. \eqref{eq:bb_transition}, we can consider the following form of $\mathbb{J}$ for simplicity,
\begin{align} \label{eq:J}
\beta \mathbb{J} = \beta \left( \begin{matrix} J_1 & & \\ & J_1 & \\ & & J_3 \end{matrix}\right) 
\end{align}
where $J_1$ specifies the coupling of the in-plane degrees of freedom and $J_3$ the coupling between the primary axes. Therefore this form of $\mathbb{J}$ is allowed for all axial groups and quantifies the anisotropy between the in-plane order and axial order, as was considered in Section \ref{sec:J_op_structure} in terms of the symmetries. 

The fact that the phase transitions are tuned with respect to the temperature $\beta=1/T$ reduces the the independent dimensionless couplings to two in terms of the reduced temperatures $\beta J_1$ and $\beta J_3$. 
Alternatively, we can consider the temperature $T$ as the tuning parameter in a thermotropic system and the anisotropy $\frac{J_1}{J_3}$ as a fixed microscopic parameter. The ratio $\frac{J_1}{J_3}$ is in fact an analogue to the so-called biaxiality parameter of $D_{2h}$ nematics \cite{SonnetVirgaDurand03, DeMatteisVirga05, DeMatteisRomanoVirga05, BiaxialBook2015}.
Accordingly, when  $\frac{J_1}{J_3}$ is sufficiently small, upon increasing temperature we expect that the in-plane order will be lost while the axial order still persist, leading to the generalized biaxial-uniaxial transition given in Eq. \eqref{eq:aixal_transition}. In the opposite limit, where $\frac{J_1}{J_3}$ is sufficiently large, it is possible to disorder the axial order while the in-plane order is still maintained, leading to the generalized biaxial-biaxial* transitions characterized by Eq. \eqref{eq:bb_transition}. Between these two limiting cases we expect direct transitions from the biaxial nematics to the $O(3)$ isotropic liquid. Note however that in general the ``biaxial" in-plane order is much more fragile than the uniaxial order of the \jaakko{primary, axial} axis. Furthermore, the biaxial in-plane order reinforces the uniaxial order since it fixes the perpendicular axial order up to a sign. Conversely, the presence of the axial order reinforces the biaxial order much less, since ordering along $\vek{n}$ still leaves in-plane $SO(2)$ fluctuations before the full ordering sets in.  As has been discovered in Ref. \cite{LiuEtAl2015b}, the highly symmetric order parameter fields experience giant fluctuations and generalized biaxial nematics with a more symmetric in-plane structure require much larger $\frac{J_1}{J_3}$ to stabilize the in-plane order.

Nevertheless, although $\frac{J_1}{J_3}$ parameterizes the anisotropy of the in-plane and axial order of general biaxial nematics, they are defined in the gauge theory, so their values do not directly indicate the relative strength of the in-plane order and axial order. Therefore $\frac{J_1}{J_3} > 1$ does not necessary mean the in-plane order is favored, and vice versa. 
Moreover, due to the $O(3)$ constraints, naturally only two of the orthonormal triads are fully independent. In the gauge theoretical effective Hamiltonian terms respecting all the symmetries and the $O(3)$ constraints, i.e. all gauge invariant combinations, appear order by order. That is, gauge invariant interactions such as $(\vek{l}_i\times \vek{m}_i) \cdot (\vek{l}_j\times \vek{m}_j) = \sigma_i\sigma_j \vek{n}_i \cdot \vek{n}_j$ or $(\vek{l}_i\cdot \vek{l}_j)^2 + (\vek{l}_i \cdot \vek{m}_j)^2 +(\vek{m}_i\cdot \vek{l}_j)^2 + (\vek{m}_i\cdot \vek{m}_j)^2 \sim (\vek{n}_i\cdot \vek{n}_j)^2$ are present with coefficients parametrized by powers of $J_1$. Therefore, eventhough $J_3 = 0$, effective axial interactions $J_{3 \rm, eff}(J_1) \sigma_i \sigma_j \vek{n}_i \cdot \vek{n}_j$ or $J'_{3,\rm eff}(J_1)(\vek{n}_i\cdot \vek{n}_j)^2$ (pseudo vector or uniaxial terms) are generated at all orders for all axial groups if allowed by the symmetries. In particular this affects higher order axial symmetries that have high rank order parameter tensors with large fluctuations. Amongst other things, due to the induced axial terms that are more relevant than the higher order in-plane interactions, the uniaxial (or uniaxial*) phase is always stabilized before the biaxial (or biaxial*) phase for in-plane symmetries with higher symmetries. The qualitative effect of these induced terms on the phase diagram is depicted in Fig. \ref{fig:J1J3}. We will see concrete examples  how these induced interactions affect the numerical phase diagrams in Section \ref{sec:MC_results}.

\subsection{Topology of the phase diagrams}\label{sec:J_phase_diagram}

\begin{figure}
\includegraphics[width=0.4\textwidth]{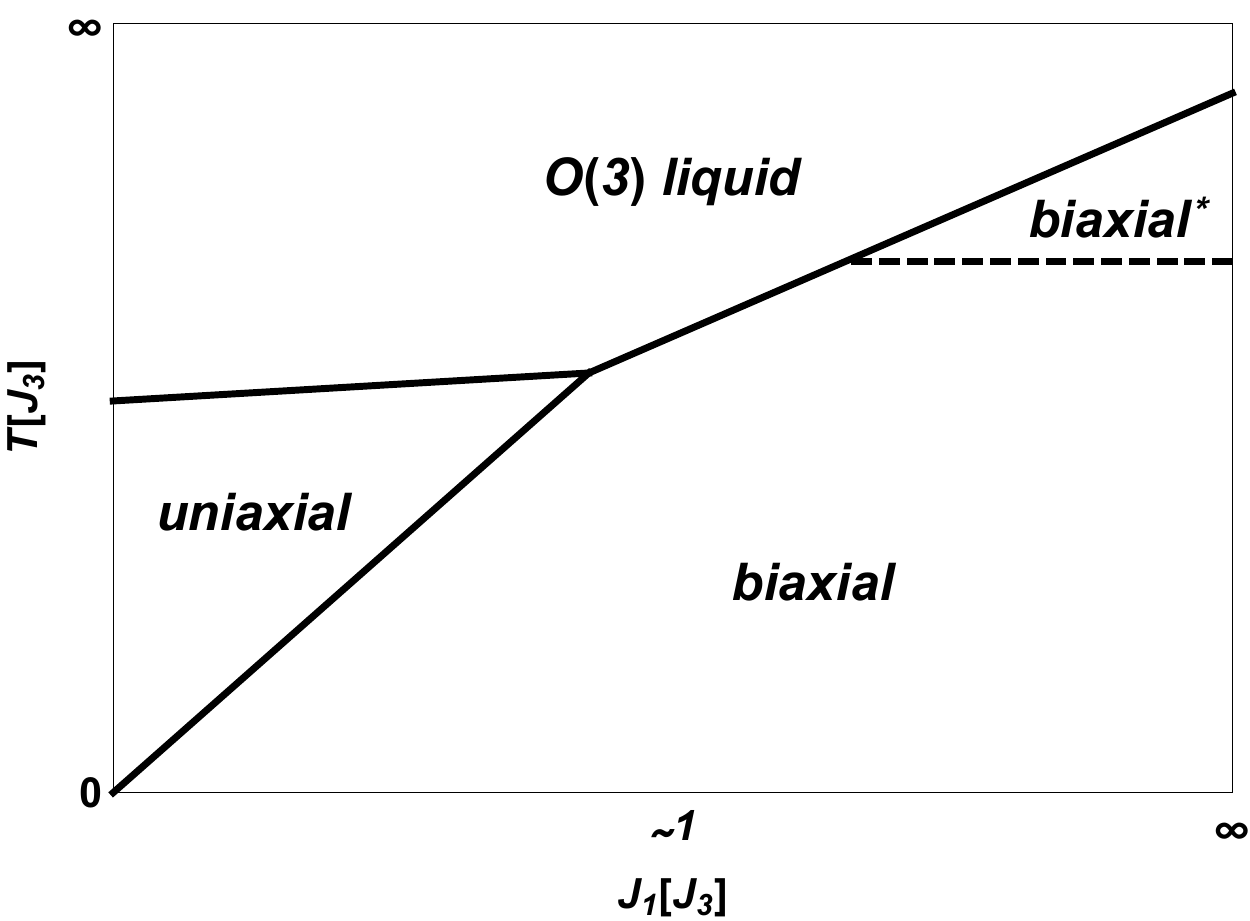}
\caption{The schematic temperature-anisotropy phase diagram of axial nematics with conventional two-fold biaxial symmetries. 
Small and large $\frac{J_1}{J_3}$ correspond to weak and strong in-plane order, respectively.
$(\frac{J_1}{J_3})^U_c$ and $(\frac{J_1}{J_3})^B_c$ are the critical anisotropies where the generalized biaxial-uniaxial transitions in Eq. \eqref{eq:aixal_transition} and the biaxial-biaxial$^*$ transitions in Eq. \eqref{eq:bb_transition} terminate, respectively.
Solid lines in the phase diagram are present for all axial symmetries $\{C_n, C_{nv}, S_{2n}, C_{nh}, D_n, D_{nh}, D_{nd} \}$ with finite $n$, while the dashed line transition is present only for the symmetries $\{C_n, C_{nv}, S_{2n}, D_n, D_{nd} \}$.
}
\label{fig:T_J1}
\end{figure}

\begin{figure}[!h]
\center
 
 \subfigure[]
 { \includegraphics[width=0.45\textwidth]{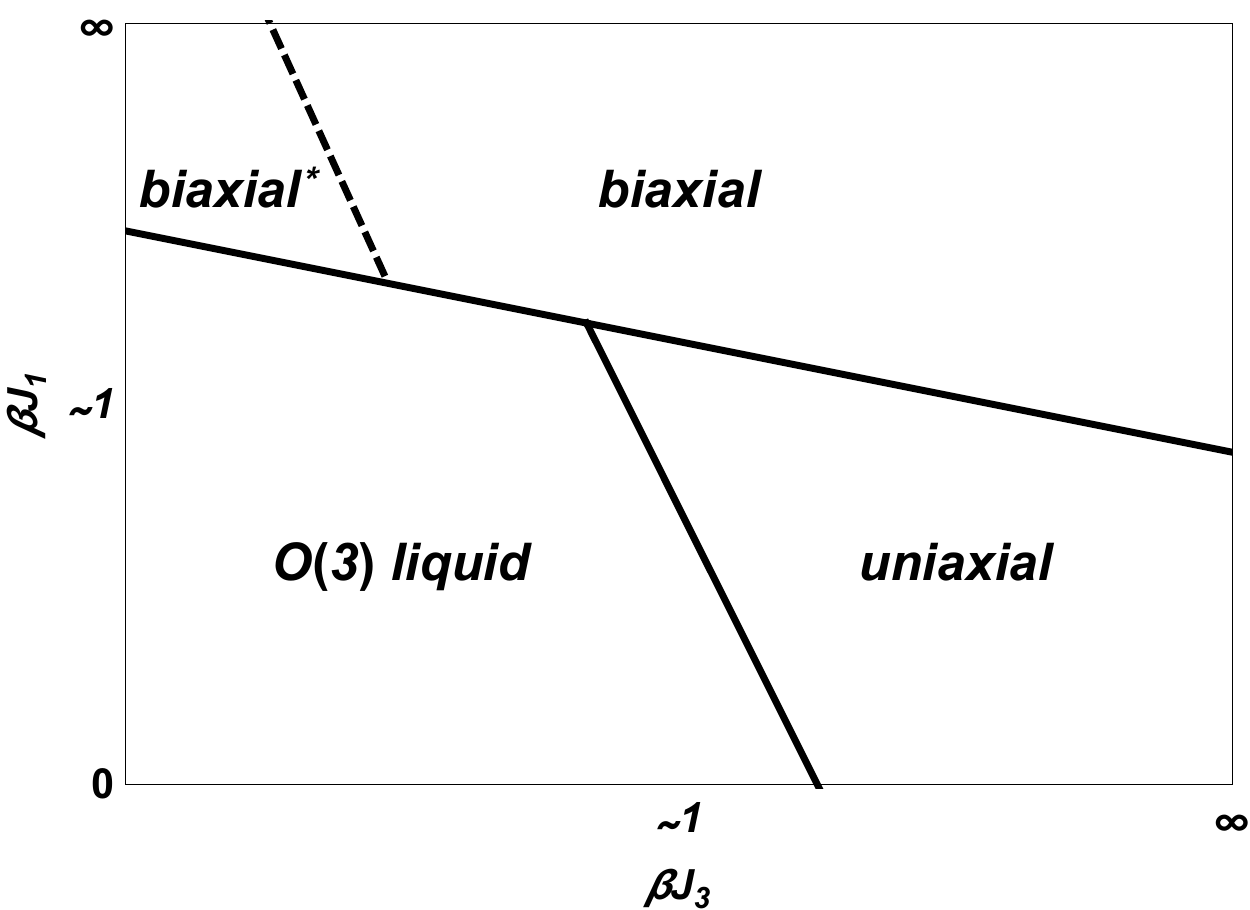} }
\vfill
 
 \subfigure[]
 { \includegraphics[width=0.45\textwidth]{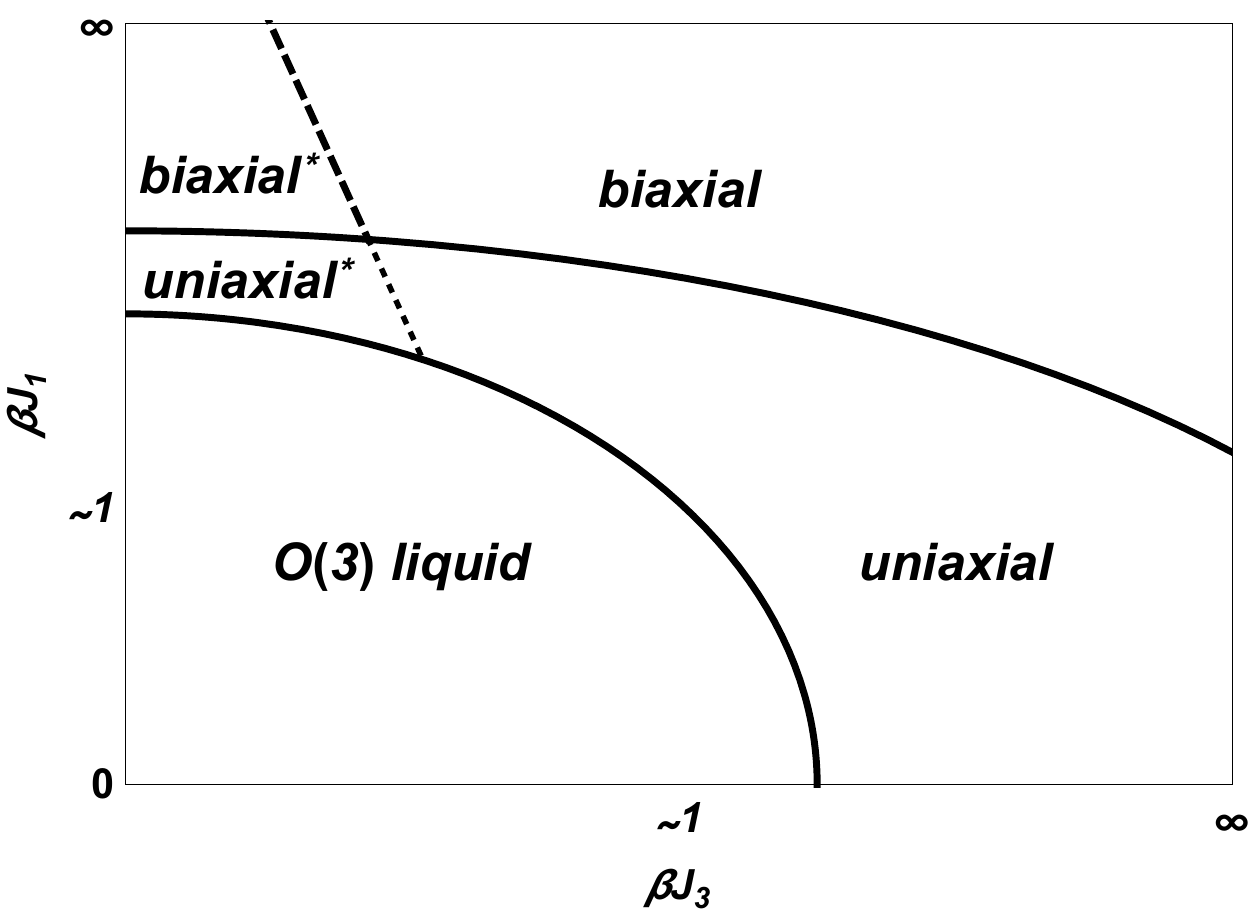} }

\caption{The schematic $(\beta J_3, \beta J_3)$ phase diagrams of axial nematics. (a): The phase diagram Fig. \ref{fig:T_J1} in terms of $(\beta J_1,\beta J_3)$. As in Fig. \ref{fig:T_J1}, for low order groups with two- and three-fold symmetries the effective couplings stabolizing the biaxial and uniaxial order are of the same order and lead to a transition directly to the biaxial phase. (b): For higher in-plane symmetries, the biaxial phase is suppressed in comparison to the uniaxial phase.  When allowed by the symmetries, axial terms with a vector or second rank uniaxial order parameter appear always in the Hamiltonian even at $J_3 = 0$ due to the $O(3)$ constraints. These always stabilize the uniaxial order while the higher order biaxial order is still fluctuating.  Solid lines in the phase diagram are present for all axial symmetries $\{C_n, C_{nv}, S_{2n}, C_{nh}, D_n, D_{nh}, D_{nd} \}$ with finite $n$, while the dashed biaxial*-transition is present only for the symmetries $\{C_n, C_{nv}, S_{2n}, D_n, D_{nd} \}$ and the dotted uniaxial*-transition for $\{C_n, C_{nv}\}$.
}
\label{fig:J1J3}
\end{figure}

Based on the discussions 
in Sections \ref{sec:biaxial_transition} and \ref{sec:J_anisotropy}, we can now identify the topology of phase diagrams of biaxial nematics at different temperatures and anisotropies of $\mathbb{J}$ as defined in Eq. \eqref{eq:J}. These are shown in Figs. \ref{fig:T_J1} and \ref{fig:J1J3}. In Fig. \ref{fig:T_J1}, we show the conventional phase diagram in terms of the temperature and the ``biaxiality" parameter $\frac{J_1}{J_3}$. In Fig. \ref{fig:J1J3} we vary the reduced axial and in-plane couplings $(\beta J_1, \beta J_3)$ independently since these relate more directly to the independent coupling strengths of the separate nematic orders in contrast to the relative anisotropy. 

Let us start with the features of the phase diagram shown in Fig. \ref{fig:T_J1} and Fig. \ref{fig:J1J3} (upper panel). As we discussed, the strength of the biaxial order should reinforce the uniaxial order more than the uniaxial order reinforces the biaxial ordering, affecting the transition temperatures. Moreover, as has been discussed in Section \ref{sec:J_anisotropy}, biaxial nematics with a more symmetric in-plane structure require larger $\beta J_1$ to stabilize the in-plane order. The critical anisotropy $(\frac{J_1}{J_3})^U_c$ for the uniaxial-biaxial transitions will therefore move to the right for biaxial nematics \jaakko{having} a larger in-plane $n$-fold rotational symmetry or more in-plane reflections. One the other hand, since a weaker in-plane order in turn means effectively stronger axial order, the critical anisotropy $(\frac{J_1}{J_3})^B_c$ for the biaxial-biaxial$^*$ transitions will correspondingly also move to the right. Therefore this phase region shrinks, while the uniaxial phase should become more prominent.

In the $(\beta J_1, \beta J_3)$-phase diagram of Fig. \ref{fig:J1J3}, the corresponding points move to the opposite directions, similarly enlarging the uniaxial region and shrinking the biaxial* phase. At the same time, as the symmetry increases, the biaxial order fluctuates more strongly leading to the the transition to the biaxial phase at considerably lower temperatures. In addition to these general trends, for higher order symmetries, the presence of the induced axial couplings rounds the phase transitions to the uniaxial phase from the isotropic liquid and leads to a finite region where only the uniaxial phase is stabilized without a direct transition to the biaxial phases. In this region, at small enough $\beta J_3$, it is possible to stabilize only the more disordered uniaxial* phase with higher $\vek{n}\to -\vek{n}$ symmetry, if the original uniaxial order is vectorial. At larger $\beta J_1$, the \jaakko{uniaxial vector order} is again lost in the biaxial*-biaxial transition. As summarized in Table \ref{table:biaxial_transitions}, the \jaakko{uniaxial* phase occurs} only for the groups $C_n, C_{nv}$. In the case of $D_n$, the uniaxial*-transition is not possible but the biaxial*-biaxial transition persist due to the non-zero chiral order parameter $\sigma$ in the $D_n$ biaxial phase, whereas the biaxial* phase has the symmetry $D_{nh}$.

Lastly, although the gauge theoretical formulation is not realized microscopically in any condensed matter system, it encodes the mesogenic symmetries very efficiently and we expect the qualitative features and the topology of the phase diagrams to be applicable to many generalized nematic systems. This is clear from the biaxial-uniaxial phase diagrams (symmetries $D_2$ and $D_{2h}$) where all expected features of the mean-field phase diagram are recovered \cite{SonnetVirgaDurand03, BiaxialBook2015}. \ke{Moreover, \josko{in }agree\josko{ment} with Ref.~\onlinecite{SonnetVirgaDurand03}, we also see evidence of a tricritical point along the biaxal-uniaxial line, as will be discussed in more detail in Section \ref{sec:MC_results}}.

\setlength{\tabcolsep}{2pt}
\renewcommand{\arraystretch}{1.6}
\begin{table*}[!tp]
\centering
\caption{{\bf A selection of three-dimensional nematic order parameters.} The first column specifies the symmetries, the second column the type $\mathbb{A}, \mathbb{B}$ of the order parameter, and the third column gives the explicit form of the order parameter tensors \cite{NissinenEtAl16}. Besides the tensors shown here, chiral nematics $D_n$ have in addition a chiral order parameter $\sigma$ defined by Eq. \eqref{eq:sigma}. } \label{table:J_ops}
\begin{tabular}{ | c | c | c | c |}
    \hline
    \hline
   \parbox{1.5cm}{\centering{\bf Symmetry \\ Groups}} 
	& \parbox{1.5cm}{\centering{\bf Type }}   
   & {\bf Ordering Tensors}  
   & \parbox{1.2cm}{\centering{\bf Tensor \\ Rank}}
   \\ \hline

   $D_2$, $D_{2h}$ 
    & $\mathbb{B}[\mathbf{l}, \mathbf{m}]$
   & $\mathbf{l} \otimes \mathbf{l} - \mathbf{m} \otimes \mathbf{m}$
   & 2
   \\ \hline
     
$D_3$, $D_{3h}$
    & $\mathbb{B}[\mathbf{l}, \mathbf{m}]$
   & $ \big( \mathbf{l}^{\otimes 3} - \mathbf{l} \otimes \mathbf{m}^{ \otimes 2} - \mathbf{m} \otimes \mathbf{l} \otimes \mathbf{m} - \mathbf{m}^{ \otimes 2} \otimes \mathbf{l}\big)$
   & 3
   \\ \hline      
     
      $D_4$, $D_{4h}$
   & $\mathbb{B}[\mathbf{l}, \mathbf{m}]$
   & \parbox{8cm}{\centering
   $\mathbf{l}^{\otimes 2} \otimes \mathbf{m} ^{\otimes 2} + \mathbf{m}^{\otimes 2} \otimes \mathbf{l} ^{\otimes 2} 
   -\frac{4}{15}  \delta_{ab} \delta_{cd} \bigotimes_{\substack{ \mu =  a,b,c, d}} \mathbf{e}_{\mu}
   + \frac{1}{15} \big(  \delta_{ac} \delta_{bd} \bigotimes_{\substack{ \mu =  a,c,b, d}} \mathbf{e}_{\mu}
    +  \delta_{ad} \delta_{bc}\bigotimes_{\substack{ \mu =  a,d,b, c}} \mathbf{e}_{\mu}  \big)$}
    & 4
   \\ \hline
     
   \parbox{3.cm}{\centering $D_{n}, D_{nh}, D_{\infty h}$}
      & $\mathbb{A}[\mathbf{n}]$
   & $ \mathbf{n} \otimes \mathbf{n} - \frac{1}{3} \id$ 
   & 2
   
   \\ \hline 
  
   \hline 
\end{tabular}

\end{table*}

\section{Quantitative phase diagrams of the gauge theoretical description} \label{sec:MC_results}

Having introduced the general concepts and framework, we still need to explicitly verify the generalized biaxial phase transitions given by Eqs. \eqref{eq:aixal_transition} and \eqref{eq:bb_transition} departing from the gauge theoretical description. 
For this purpose we have simulated the temperature-anisotropy phase diagram and the $J_1$-$J_3$ phase diagrams for various symmetries, using the standard Metropolis Monte-Carlo algorithm. These simulations were performed on lattices having dimensions $L^3 = 8^3, 10^3, 12^3, 16^3$. 
The associated order parameters and their characterizations relevant for the phase transitions are collected in Table \ref{table:J_ops} and Table \ref{table:biaxial_transitions}, respectively. 
As we detail below, the obtained results completely agree with the general scenario of generalized biaxial phase transitions \jaakko{as discussed in the previous sections}.

\subsection{Determination of the phases}

To determine the symmetry of a nematic phase with tensor order parameter $\mathbb{O}^{G}$, one in principle needs to consider all the entries of $\mathbb{O}^{G}$. 
However, for interactions favoring homogeneous distribution of the order parameter fields, such as the interaction in the gauge model Eq. \eqref{eq:J_gauge_theory}, the symmetry of the phase can be revealed by the strength of the order parameter defined as 
\begin{align}
q = \sqrt{ (\mathbb{O}^G_{abc...})^2},
\end{align}
where $\mathbb{O}^G= \frac{1}{L^3} \sum_i \mathbb{O}^{G}_i$, averages the order parameter tensor over the system, $a, b, c, \dots$ denote the tensor components and contractions for repeated tensor indices are assumed. In combination with symmetry arguments, the scalar order parameter is enough to fix the symmetry of the phase and the nematic ordering strength will develop a finite value in the ordered phase and vanish in the disordered phase  (For more details, see, e.g., Refs. \cite{LiuEtAl2015b, NissinenEtAl16}.).

For axial nematics, we accordingly need to define the ordering strength for the axial order $\mathbb{A}^G$ and the in-plane order  $\mathbb{B}^G$, respectively,
\begin{align}
q_A = \sqrt{ (\mathbb{A}^G_{ab...})^2},
\\
q_B = \sqrt{ (\mathbb{B}^G_{ab...})^2}.
\end{align}
A transition is then identified by monitoring the appearance of a peak in the associated susceptibility
\begin{align}
\chi(q_{A,B}) = \frac{L^3}{T}(\langle q^2_{A,B} \rangle - \langle q_{A,B} \rangle^2).
\end{align}
where $\langle ... \rangle$ denotes the thermal average

Moreover, we have also computed the heat capacity and the susceptibility of the chiral order parameter, which are defined in the usual way,
\begin{align}
C_v &= \frac{1}{T^2 L^3} (\langle E^2 \rangle - \langle E \rangle^2),
\\
\chi(\sigma) &= \frac{L^3}{T}(\langle \sigma^2 \rangle - \langle \sigma \rangle^2).
\end{align}
where $E$ is the internal energy of the system, and $ \sigma = \frac{1}{L^3} \sum_i \sigma_i$ is the global chiral order parameter.

\begin{figure}
\centering
 \subfigure[]{
  \includegraphics[width=0.45\textwidth]{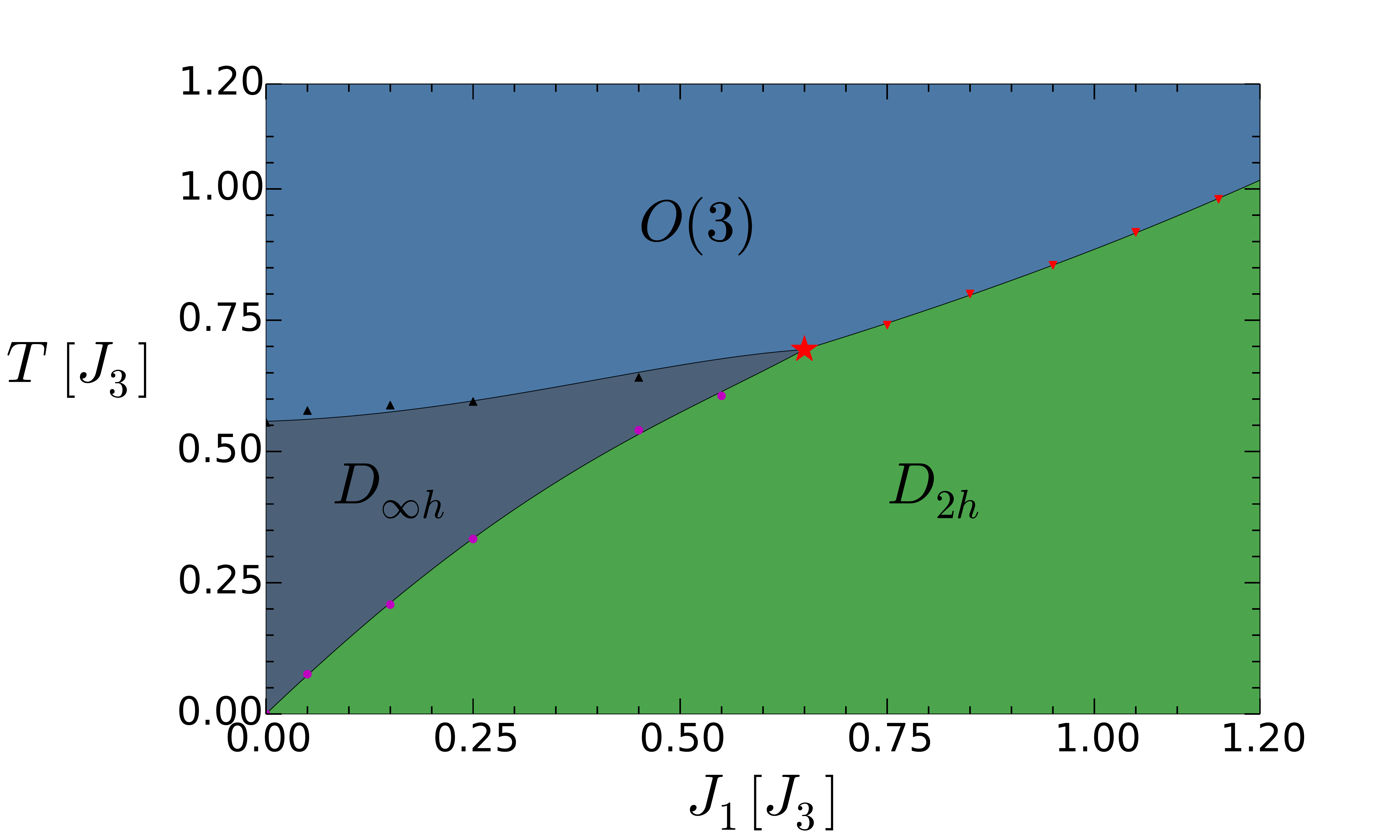}\label{fig:D2h_J1}
 }
 \vfill
 \subfigure[]{
  \includegraphics[width=0.45\textwidth]{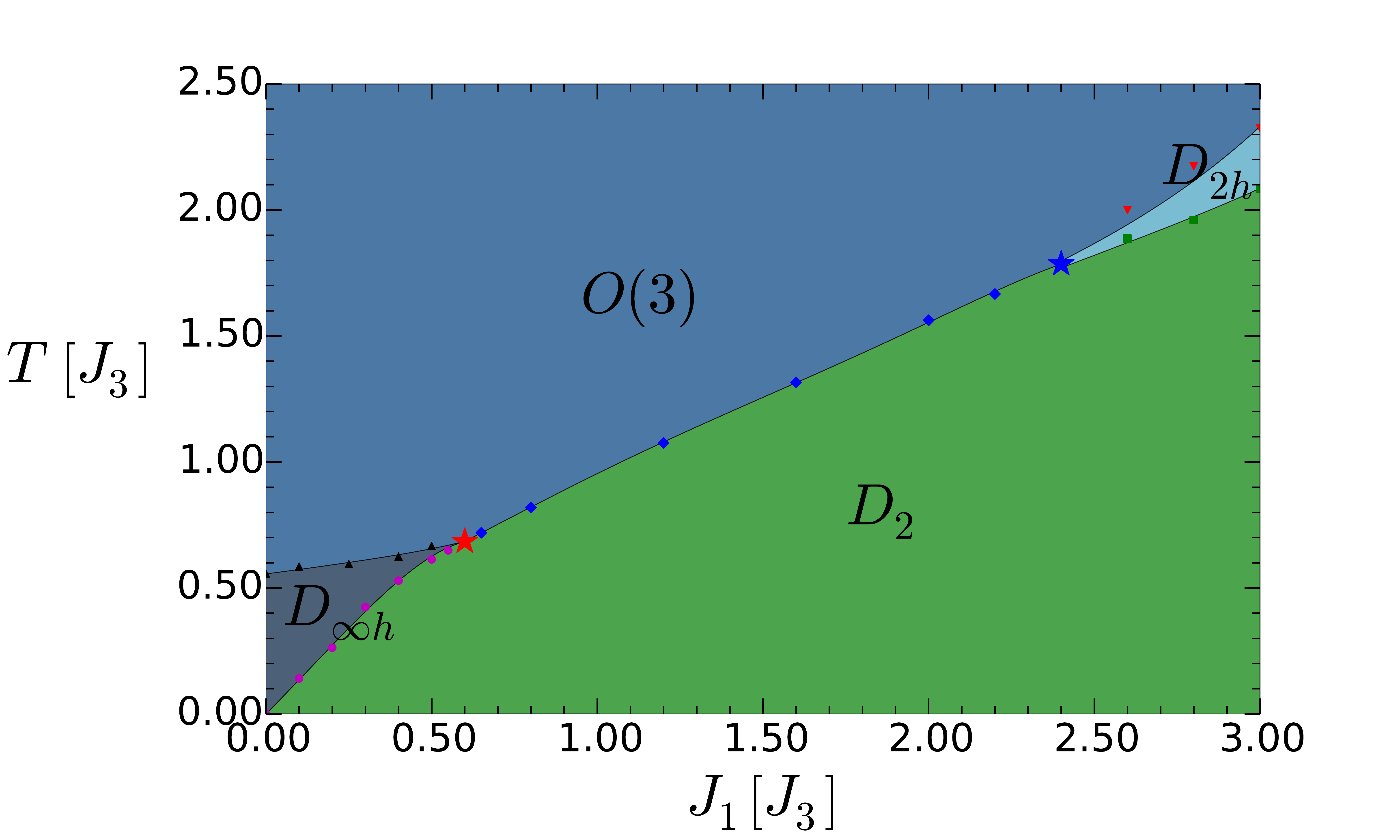}\label{fig:D2_J1}
 }
 
 \caption{ The temperature-anisotropy phase diagram of \subref{fig:D2h_J1} $D_{2h}$ and \subref{fig:D2_J1} $D_2$ biaxial nematics. At small $\frac{J_1}{J_3}$, there is a sequence of biaixal-uniaxial-liquid transition with a vestigial $D_{\infty}$-uniaxial phase. 
 The uniaxial phase terminates at a triple point (the red star), after which the transition sequence truns to a direct biaxial-liquid transition.
 This directly reproduces the well-known phase transitions for $D_{2h}$ and $D_2$ nematics from the gauge theoretical setup \eqref{eq:J_gauge_theory}.
In addition, for large $\frac{J_1}{J_3}$ in the $D_2$ case, there is a vestigial $D_{2h}$-biaxial phase right to another triple point (the blue star), realizing the biaxial-biaxial$^*$ transition in Eq.~\eqref{eq:bb_transition}. 
   \label{fig:D2h_D2_J1}}
\end{figure}

\subsection{Phase diagrams involving temperature versus anisotropy } 
\jaakko{A salient feature of our results is} that we can retrieve the well-known temperature-anisotropy phase diagram of  $D_{2h}$ and $D_2$ nematics within our gauge theoretical setting \eqref{eq:J_gauge_theory}, see Fig. \ref{fig:D2h_D2_J1}.
In the region of small $\frac{J_1}{J_3}$, where the stiffness of the in-plane order is weaker than that of the axial order, we see that upon increasing the temperature, the in-plane order is destroyed first, leaving room for a vestigial $D_{\infty h}$-uniaxial phase.
This vestigial uniaxial phase vanishes at a critical anisotropy $(\frac{J_1}{J_3})^U_c$, after which the in-plane and axial order become of comparable strength and the transitions merges into a single transition between the biaxial phase and the $O(3)$ liquid phase.
On the other hand, for large $\frac{J_1}{J_3}$, when the in-plane order is sufficiently strong, there will be a vestigial $D_{2h}$-biaxial phase in the $D_2$ case Fig.~\ref{fig:D2_J1}.
This realizes the biaxial-biaxial$^{*}$ transition in Eq. \eqref{eq:bb_transition}.
We note however that, as discussed in Section \ref{sec:J_anisotropy}, although the in-plane coupling can effectively induce an axial coupling, the axial order is not fully destroyed during this transition.
The resulting behavior of the associated order parameters across these transitions are given in Table \ref{table:biaxial_transitions}.

Moreover, we find that the direct transition between the $D_2$-biaxial nematic phase and the $O(3)$ isotropic liquid phase  in Fig.~\ref{fig:D2_J1} is first-order like.
Both $\chi(q_B)$, $\chi(\sigma)$ and $C_v$ exhibit a sudden peak at the transition, and the magnitude of their peak grows dramatically with the lattice size.
This discontinuity continues to the biaxial-uniaxial transition line.
Therefore, we identify a triple point \jaakko{where the three transition lines in Fig. \ref{fig:D2_J1} meet \josko{and} the three phases can coexist}.
Moreover,  in the middle of the biaxial-uniaxial transition line we find evidence for a tricritical point where the first order phase transition terminates and the transition becomes continuous. \jaakko{These observations {\it exactly agree with the mean field phase diagram and the experimental results} of biaxial nematics in Refs. \onlinecite{SonnetVirgaDurand03,MerkelEtAl04}}.

Besides these two familiar examples, we have also verified the generalized uniaxial-biaxial transitions in Eq. \eqref{eq:aixal_transition} as well as the uniaxial-uniaxial* and biaxial-biaxial$^*$ transition in Eq. \eqref{eq:bb_transition} for nematics having symmetry
$\{S_2, C_2, C_{2v}, C_{2h}, D_{2d}, S_4, D_{3}, D_{3h}, C_{4v}, D_4, D_{4h}, D_6, D_{6h} \}$.
These comprise all the seven types of axial groups and include symmetries with low and high symmetric in-plane structure.
For nematics with a low symmetry, including the cases $\{S_2, C_2, C_{2v}, C_{2h}, D_{2d}, S_4, D_{3}, D_{3h} \}$, the phase diagrams have been checked to have the same topology as those of the $D_2$ or $D_{2h}$ case and are thus  not presented here.
For nematics with a high symmetry, comprising the cases $\{ D_4, D_{4h}, D_6, D_{6h} \}$, the generalized biaxial transitions will however be affected dramatically by the induced axial coupling, as discussed in Section \ref{sec:J_phase_diagram}, and the phase diagrams are different.
In next section, we will discuss these phase diagrams for each of these symmetries.

\begin{figure}[!tp]
\centering
 \subfigure[]
 {
  \includegraphics[width=0.45\textwidth]{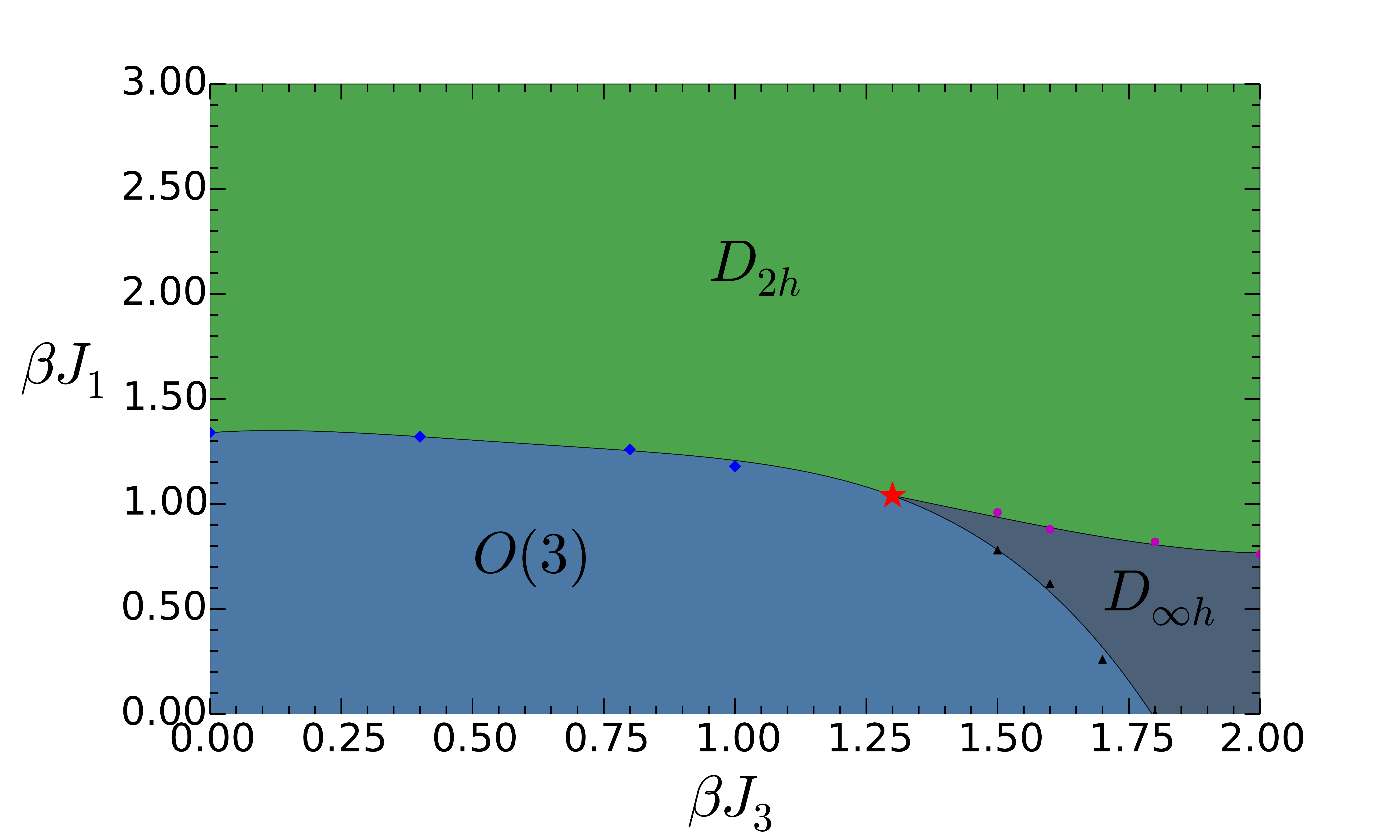}\label{fig:D2h_J13}
 }
 \vfill
 \subfigure[]
 {
  \includegraphics[width=0.45\textwidth]{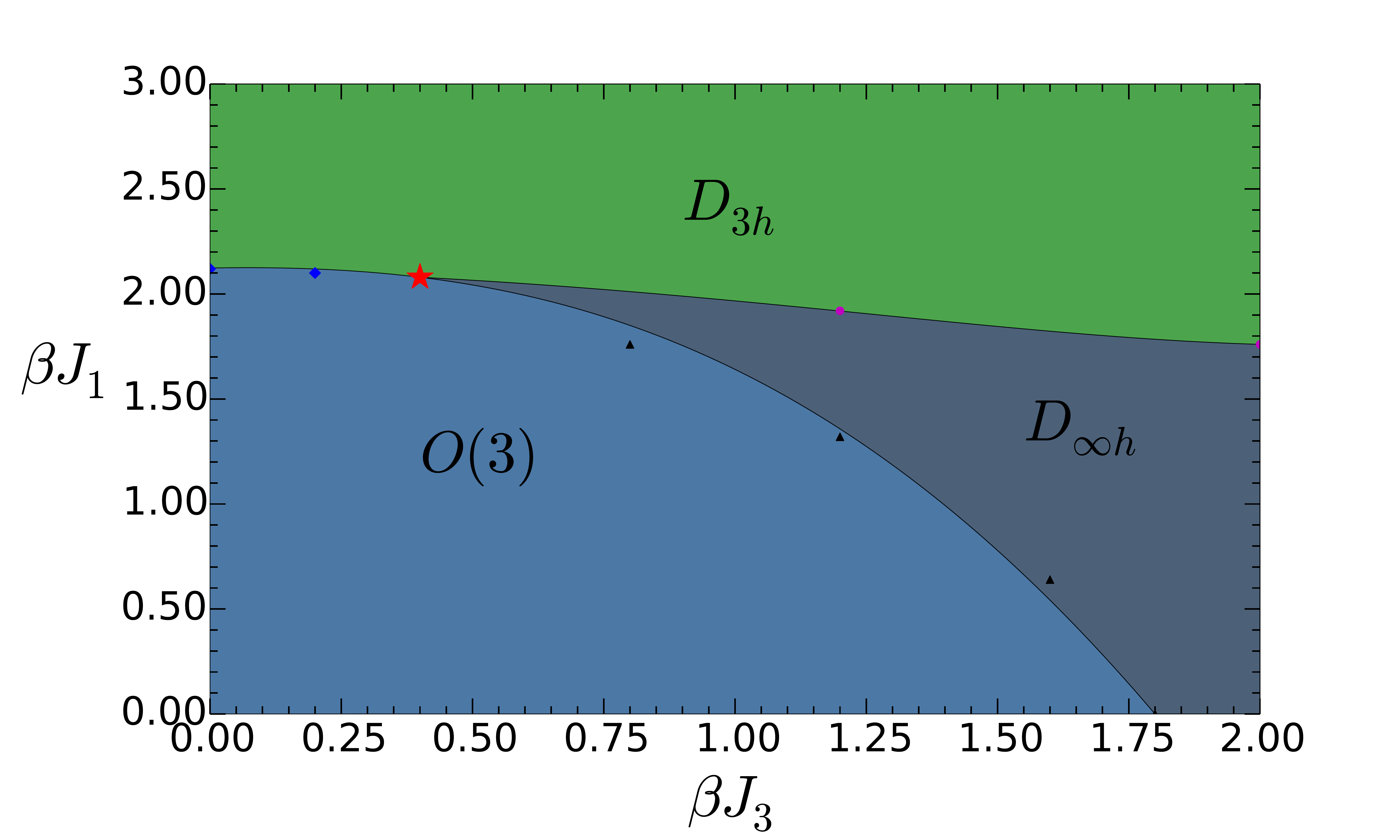}\label{fig:D3h_J13}
  }
  \vfill 
 \subfigure[]
 {
  \includegraphics[width=0.45\textwidth]{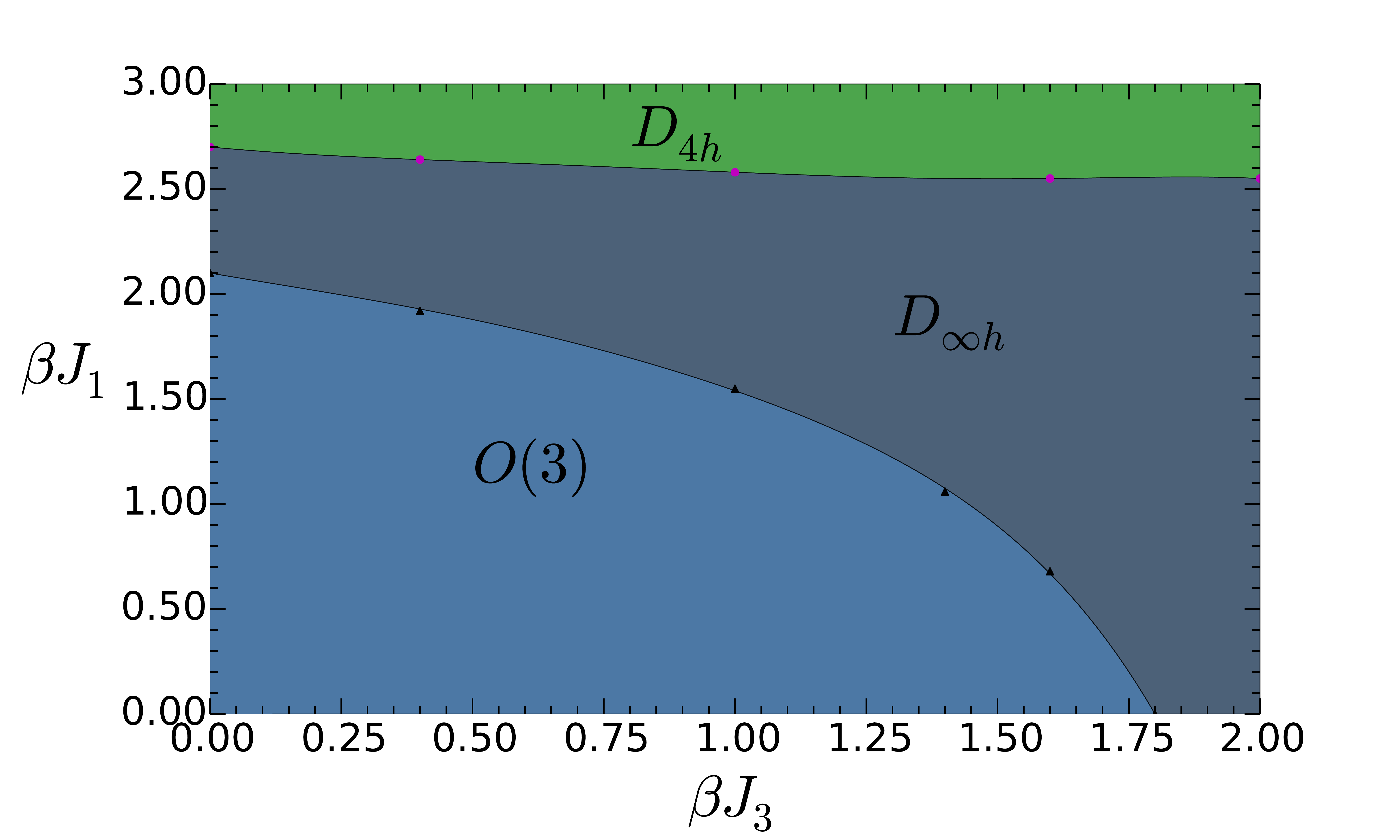}\label{fig:D4h_J13}
 }
 
 \caption{ The $J_1$-$J_3$ phase diagram of \subref{fig:D2h_J13} $D_{2h}$, \subref{fig:D3h_J13} $D_{3h}$ and \subref{fig:D4h_J13} $D_{4h}$ nematics. \josko{
The red stars in \subref{fig:D2h_J13} $D_{2h}$ and \subref{fig:D3h_J13} $D_{3h}$ highlight a triple point, in which the three transition lines meet. }
Similar to the temperature-anisotropy phase diagram in Fig. \ref{fig:D2h_D2_J1}, there is a vestigial uniaxial phase appearing from the region with small $J_1$ and large $J_3$ (small $\frac{J_1}{J_3}$), realizing the generalized biaxial-uniaxial transition in Eq. \eqref{eq:aixal_transition}.
As the symmetry increases, this vestigial uniaxial phase becomes more prominent and the fully ordered biaxial phase is remarkably squeezed.
When the symmetry is sufficiently high, the vestigial uniaxial phase appears adjacent to the isotropic liquid due to the symmetry allowed axial terms.
\ke{Moreover, our simulations indicate that depending on strength of the in-plane coupling, the biaxial-uniaxial transition may be either first order or second order. 
Therefore a tricritical point may exist in the biaxial-uniaxial transition line.}
  \label{fig:D234h_J13}}
\end{figure}

\begin{figure}[!tp]
\centering
 \subfigure[]
 {
  \includegraphics[width=0.45\textwidth]{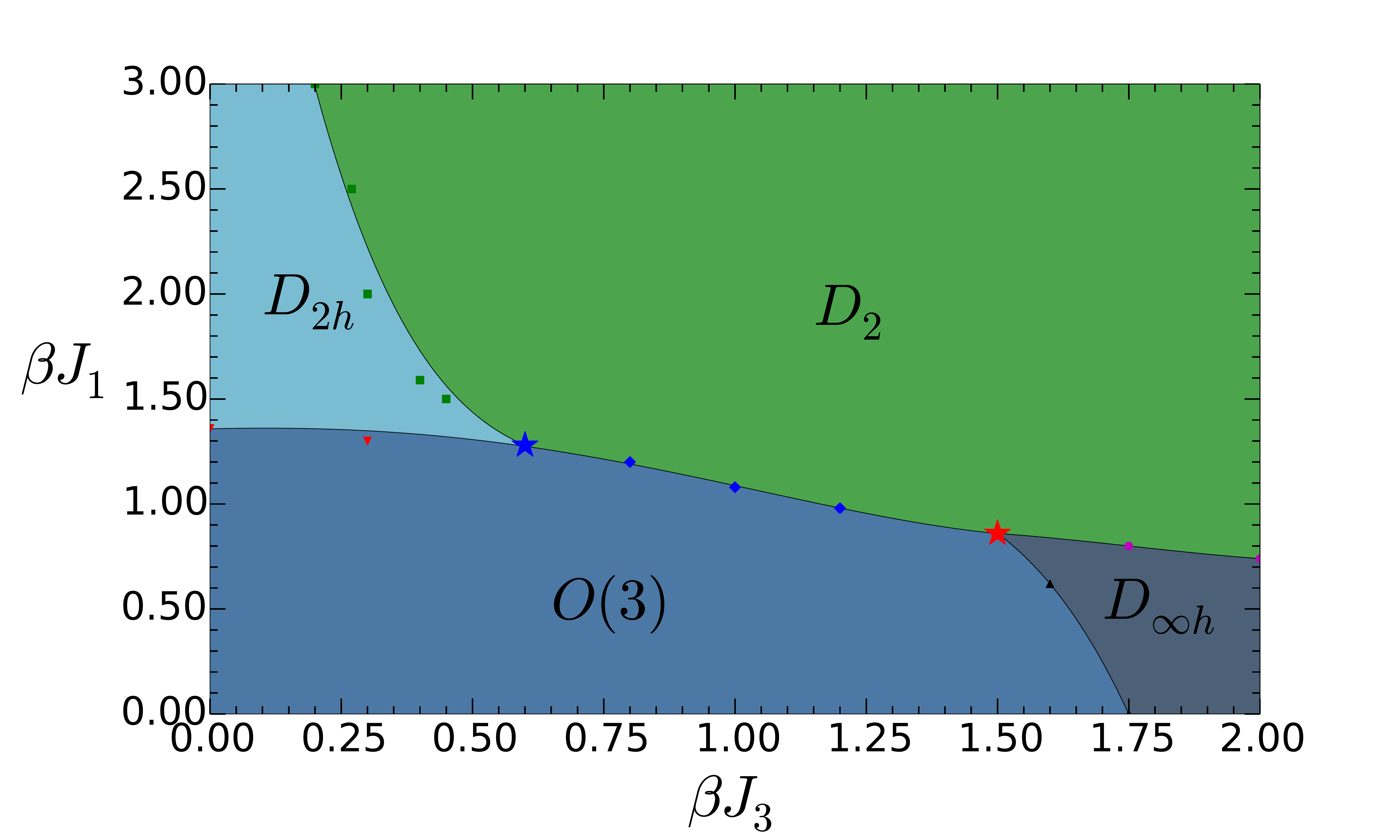}\label{fig:D2_J13}
 }
 \vfill
 \subfigure[]
 {
  \includegraphics[width=0.45\textwidth]{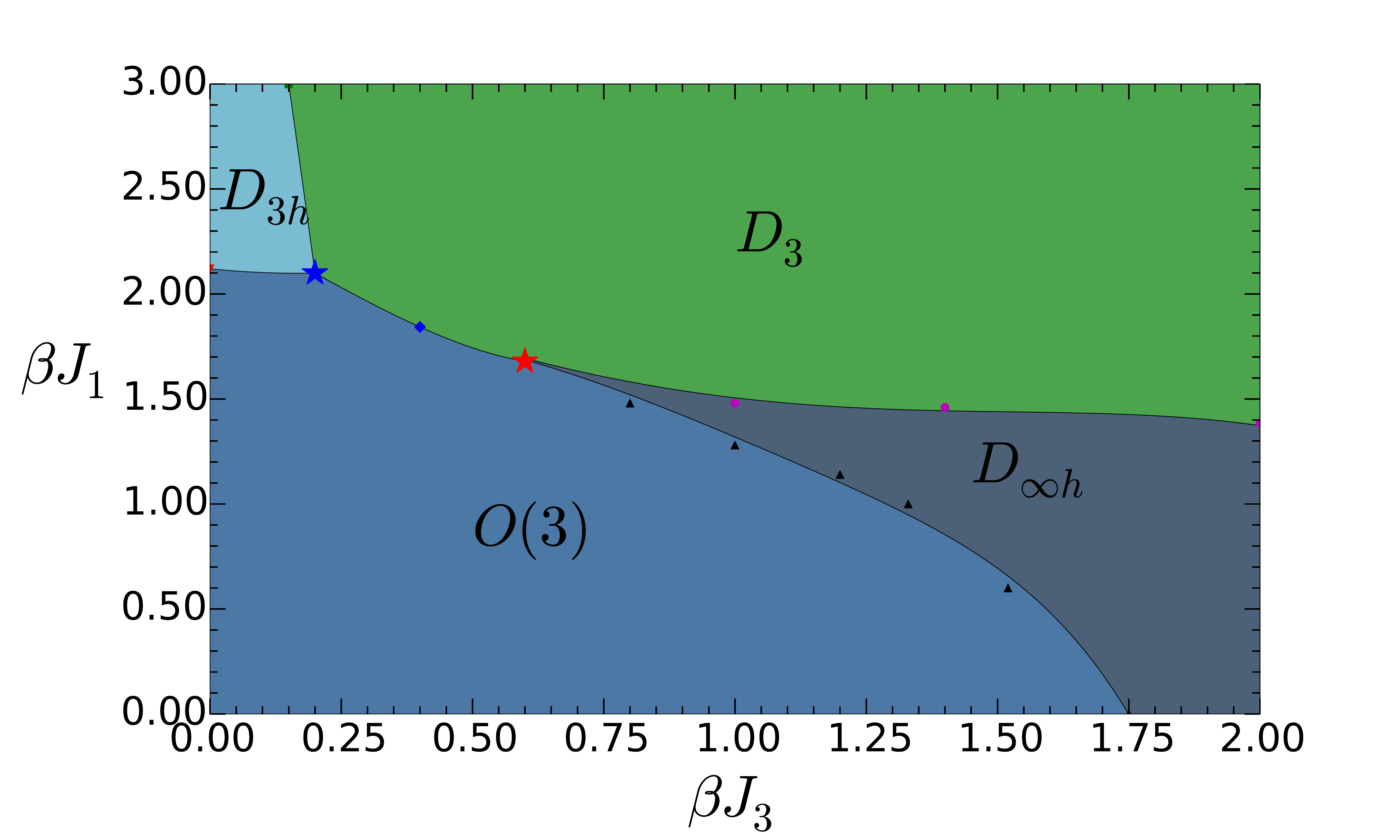}\label{fig:D3_J13}
  }
  \vfill 
 \subfigure[]
 {
  \includegraphics[width=0.45\textwidth]{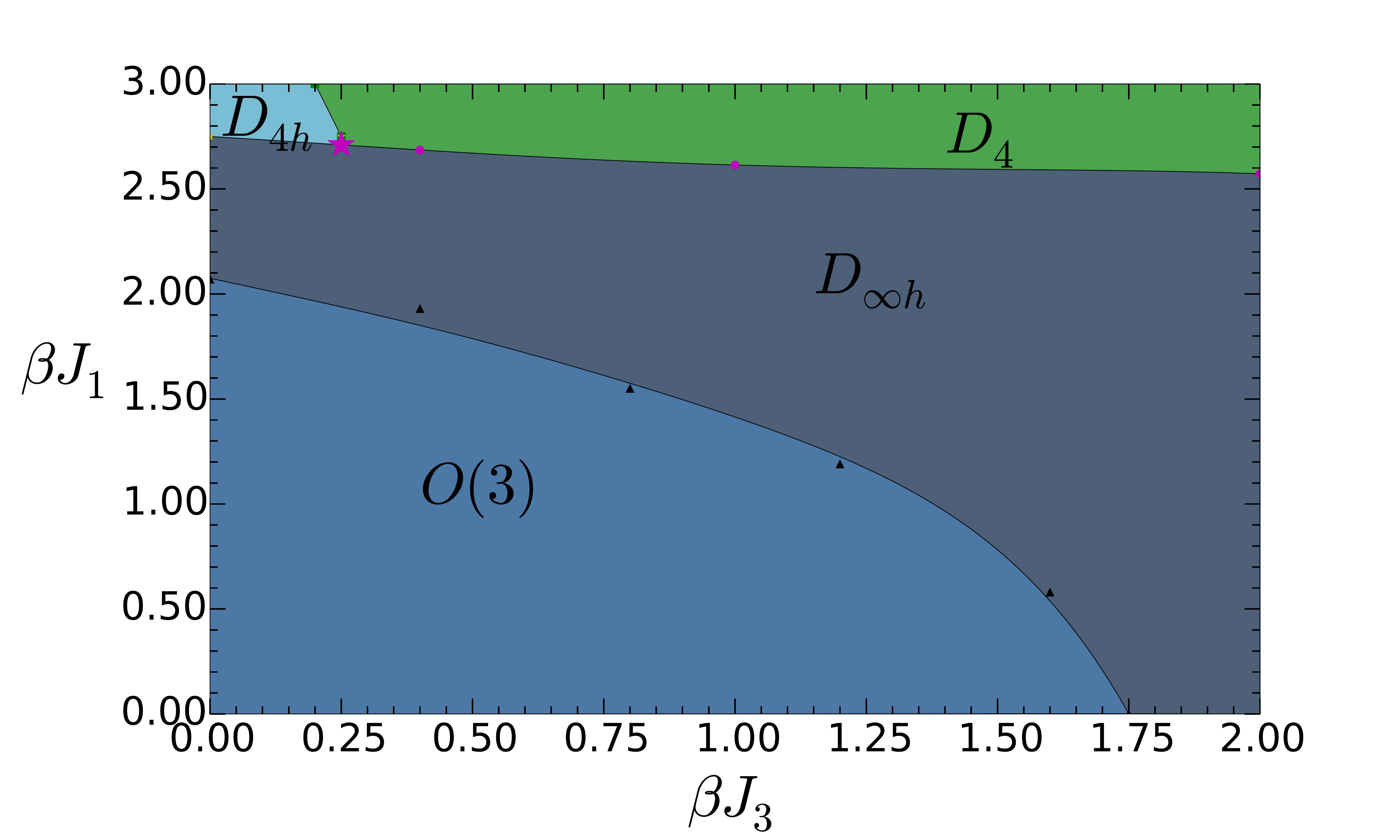}\label{fig:D4_J13}
 }
 
 \caption{ The $J_1$-$J_3$ phase diagram of \subref{fig:D2_J13} $D_{2}$, \subref{fig:D3_J13} $D_{3}$  and \subref{fig:D4_J13}  $D_{4}$ nematics. 
Similar to Fig. \ref{fig:D234h_J13}, but there is in addition a vestigial biaxial phase at small $J_3$ region, realizing the generalized biaxial-biaxial$^*$ transition in Eq. \eqref{eq:bb_transition}.
Both this vestigial biaxial phase and the fully ordered biaxial phase are squeezed considerably as the symmetry increases.
The associated triple points at where transition lines meet are highlighted by large stars. 
\ke{As in Fig. \ref{fig:D234h_J13}, there may be a tricritical point in the biaxial-uniaxial transition line.}
 \label{fig:D234_J13}}
\end{figure}

\subsection{$J_1$-$J_3$ phase diagrams} 

As already discussed in the introduction, within the gauge theoretical description we can in fact compare the physics of nematics with different symmetries in a common reference.
In Fig. \ref{fig:D234h_J13}, we show the $J_1$-$J_3$ phase diagram for $D_{2h}$, $D_{3h}$ and $D_{4h}$ nematics.
Let us first focus on the $D_{2h}$ case in Fig. \ref{fig:D2h_J13}. As in the temperature-anisotropy phase diagram in Fig. \ref{fig:D2h_J1}, in the region with small $J_1$ and large $J_3$ there is a vestigial uniaxial phase sandwiched between the fully ordered biaxial phase and the disordered liquid phase.
The critical anisotropy where the vestigial uniaxial phase starts appearing is consistent with that of Fig. \ref{fig:D2h_J1}, up to our numerical accuracy.
Moving to $D_{3h}$ case, the increased in-plane symmetry requires a larger in-plane coupling (lower temperature and larger $\frac{J_1}{J_3}$ anisotropy) to stabilize the biaxial order, due to the more severe fluctuations.
The biaxial phase is therefore squeezed by the liquid phase and the vestigial uniaxial phase.

The squeezing of the biaxial phase is even more prominent for the $D_{4h}$ nematics, where the in-plane symmetry is increased further.
In particular, since very large in-plane coupling is required to stabilize the highly symmetric $D_{4h}$ order, before the biaxial phase is realized, the induced axial coupling is always sufficiently strong for the uniaxial order.
This leads to a vestigial uniaxial phase realized for all non-negative values of the ``bare'' axial coupling $J_3$, while the direct biaxial-liquid transition is absent.
The same is true for the more symmetric $D_{6h}$ nematics, with a even larger region of the vestigial uniaxial phase.

However, one should not interpret this as a no-go theorem for a direct biaxial-liquid transition in the case of highly symmetric biaxial nematics.
Instead, this simply means that in order to realize this transition, one needs to consider a model with ``anti-nematic'' coupling for the axial order to offset the induced axial coupling.

The above discussions can similarly be verified for $D_{2}$, $D_{3}$ and $D_{4}$ nematics as well, as shown in Fig. \ref{fig:D234_J13}.
Nonetheless, since the biaxial-biaxial$^*$ transition is possible for these cases, in the small $J_3$ region, there is \jaakko{in addition a vestigial biaxial* phase}.
This phase is also squeezed as symmetries increase, as in the case of the fully ordered biaxial phase.
Moreover, in cases of $D_2$ and $D_3$, there are direct transitions from the fully ordered biaxial phase or vestigial biaxial phase to the liquid phase.
For the highly symmetric $D_4$ case, however, these transitions are replaced by a biaxial-uniaxial or a biaxial$^*$-uniaxial transition, since a vestigial uniaxial phase exists for all non-negative values of $J_3$ as in the $D_{4h}$ case \jaakko{due to the induced axial couplings}.

\section{Conclusions and outlook}\label{sec:conclusions}

There is a rich landscape of unexplored generalized nematics, entailing not only a diversity \jaakko{of orientational phases} in terms of their symmetry but also an abundance in possible vestigial phases.
In this paper, we have discussed the anisotropy-induced vestigial \jaakko{uniaxial and biaxial phases} for nematics characterized by axial point-group symmetries and studied their phase transitions.
Our results generalize the well-studied biaxial-uniaxial transition of $D_{2h}$ nematics to a much broader class, that can be directly accessed within our earlier proposed gauge theoretical formulation of generalized nematics \cite{LiuEtAl2015b} and follow \jaakko{from} a-priori symmetry arguments.
This framework allows us in particular to compare nematics \jaakko{and vestigial phases} with different symmetries in one common reference.
Utilizing this formalism, we found that, in comparison to the familiar $D_{2h}$ biaxial nematic phase, nematic phases with high axial symmetries require much lower temperature to stabilize their order.
This motivates the  fact that biaxial phases with high symmetry are difficult to realize in reality and have not yet been experimentally encountered: 
before reaching the low temperature demanded by the biaxial order, crystallization may already start playing a role.
Consequently, columnar, smectic and/or crystalline phases may occur instead of a generalized nematic phase. We stress that such states are not captured by our model that by construction encompasses only the orientational ordering. These challenges not withstanding, the advances in the fabrication and manipulation of colloidal systems of nanoparticles appear in fact promising with regards to stabilizing generalized nematic phases in the laboratory in the near future \cite{Glotzer2007, LiJosephsonStein2011, DamascenoGlotzer2012, Mark2013, Manoharan2015}.

Besides these generalized biaxial transitions, there may be more vestigial phases and transitions in the gauge model Eq. \eqref{eq:J_gauge_theory}.
Those phases are associated with the defects in the model, which have been ignored in this work by setting $H_{\rm gauge} = 0$ in Eq. \eqref{eq:J_gauge_theory}, describing the confined and Higgs phases of the model. From the point of view of topological melting, phase transitions may be understood as a proliferation of topological defects \cite{LammertRoksharToner95, LiuEtAl15, Beekman2016}. To illustrate this further we can take the $D_{2h}$-biaxial nematic as an example.
According to homotopy theory, topological defects of $D_{2h}$ nematics are classified by the five conjugacy classes of the quaternion group $Q_8$ \cite{VolovikMineev77, Mermin1979, Michel1980, XuLudwig2012}.
Among these defects, there are only three elementary ones, which are the $\pi$-disclinations in the three orthogonal planes of the three dimensional space.
In the transition of the nematic phase to the $O(3)$ liquid phase, all these defects proliferate.
In the biaxial-uniaxial transition, however, one of them stays gapped.
This implies that a phase transition can be affected by the tuning of the energy cost of topological defects.
The gauge model Eq. \eqref{eq:J_gauge_theory} provides a natural way to do this.
Concretely, when the $H_{\rm gauge}$ term is set to be zero, topological defects in the model only cost elastic energy by the $H_{\rm Higgs}$ term.
By tuning on the $H_{\rm gauge}$ term, however, we can introduce a finite core energy to a particular class of topological defects, and therefore modify the nature of the phase transition. While such defect terms $H_{\rm gauge} \neq 0$ have been identified to be important in the melting of many quantum nematics \cite{SenthilFisher2000, SenthilFisher2001, ZaanenNussinov02, ZaanenNussinov03, PodolskyDemler05, MrossSenthil12}, they have not yet been discovered to play a role in the realm of classical nematics and melting \cite{TonerLammertRokshar95b}. The rich physics associated with these ideas leave many interesting avenues of for future research in the generalized nematic systems.

\textbf{Acknowledgments} 
 This work has been supported by the Dutch Foundation on the Research of Fundamental Matter (FOM), which is part of NWO. K. L. is supported by the State Scholarship Fund program organized by China Scholarship Council (CSC).

\bibliographystyle{apsrev4-1}
\bibliography{nematics}

\end{document}